\RequirePackage{fix-cm}
\documentclass[smallextended]{svjour3}       
\smartqed  
\usepackage{graphicx}
\def\msun{M$_\odot$}
\def\iso#1{$^{#1}$}
\usepackage[numbers]{natbib}
\usepackage{url} 
\usepackage{hyperref}
\usepackage[dvipsnames]{xcolor}
\usepackage[normalem]{ulem}

\newcommand{\aap}{{\it A\&A}}
\newcommand{\apj}{{\it ApJ}}
\newcommand{\apjl}{{\it ApJL}}
\newcommand{\apjs}{{\it ApJ Suppl. Ser.}}
\newcommand{\mnras}{{\it MNRAS}}

\newcommand{\araa}{{\it Ann. Rev. Astron. Astrophys.}}

\newcommand{\prc}{{\it Phys. Rev. C}}
\newcommand{\pasa}{{\it Publ. Astron. Soc. Austr.}}
\newcommand{\gca}{{\it Geochim. Cosmochim. Acta}}

\newcommand{\ssr}{{\it Space Science Reviews}}

\begin{document}

\authorrunning{Lugaro et al.}
\titlerunning{Representation of $s$-process abundances}

\title{Representation of $s$-process abundances for comparison to data from bulk meteorites}




\author{Maria Lugaro$^{1,2,3,4}$, Mattias Ek$^{5}$, M\'aria Pet\H{o}$^{1,2}$, Marco Pignatari$^{1,2,6}$, Georgy V. Makhatadze$^{7}$, Isaac J. Onyett$^{7}$, Maria Sch\"onb\"achler$^{5}$}



\institute{
$^{1}$Konkoly Observatory, Research Centre for Astronomy and Earth Sciences, Eötvös Loránd Research Network (ELKH), H-1121 Budapest, Konkoly Thege M. \'ut 15-17, Hungary \\ 
           $^{2}$CSFK, MTA Centre of Excellence, Budapest, Konkoly Thege Miklós út 15-17, H-1121, Hungary \\ 
           $^{3}$ELTE E\"{o}tv\"{o}s Lor\'and University, Institute of Physics, Budapest 1117, P\'azm\'any P\'eter s\'et\'any 1/A, Hungary\\
          $^{4}$School of Physics and Astronomy, Monash University, VIC 3800, Australia\\ 
           $^{5}$Institute for Geochemistry and Petrology, ETH Zürich, Zurich, Switzerland\\ 
           $^{6}$E. A. Milne Centre for Astrophysics, University of Hull, Cottingham Road, Kingston upon Hull, HU6 7RX\\ 
$^{7}$Centre for Star and Planet Formation (StarPlan), Globe Institute, Faculty of Health and Medical Sciences, University of Copenhagen, \O{}ster Voldgade 5-7, 1350 Copenhagen K, Denmark
}

\date{Received: date / Accepted: date}

\maketitle

\begin{abstract}
Analysis of bulk meteorite compositions has revealed small isotopic variations due to the presence of material (e.g., stardust) that preserved the signature of nuclear reactions occurring in specific stellar sites. The interpretation of such anomalies provides evidence for the environment of the birth of the Sun, its accretion process, the evolution of the solar proto-planetary disk, and the formation of the planets. A crucial element of such interpretation is the comparison of the observed anomalies to predictions from models of stellar nucleosynthesis. To date, however, this comparison has been limited to a handful of model predictions. This is mostly because the calculated stellar abundances need to be transformed into a specific representation, which nuclear astrophysicists and stellar nucleosynthesis researchers are not familiar with. Here, we show in detail that this representation is needed to account for mass fractionation effects in meteorite data that can be generated both in nature and during instrumental analysis. We explain the required internal normalisation to a selected isotopic ratio, describe the motivations behind such representation more widely, and provide the tools to perform the calculations. Then, we present some examples considering two elements produced by the $slow$ neutron-capture ($s$) process: Sr and Mo. We show which specific representations for the Sr isotopic composition calculated by $s$-process models better disentangle the nucleosynthetic signatures from stars of different metallicity. For Mo, the comparison between data and models is improved due to a recent re-analysis of the \iso{95}Mo neutron-capture cross section.

\end{abstract}
%
%
\section{Introduction}
\label{sec:intro}
{\it 
}

Since its beginning, the study of the $slow$ neutron-capture ($s$) process has relied on the analysis of meteorites. Thanks to the derivation of the solar abundances of the $s$-only isotopes\footnote{These are isotopes that can only be produced by the $s$ process, as they are shielded from $r$-process production by a stable isotope with the same mass but lower atomic number.} from analysis of meteorites \cite{suess56} the early $s$-process studies of Don Clayton and collaborators \cite{seeger65,clayton61,clayton74} and Franz K\"appeler and collaborators \cite{kaeppeler82} were able to define the first $s$-process distributions and identify the weak, main, and strong components. The reproduction of these components allowed this pioneering research to study in great detail the characteristics of the $s$-process neutron fluxes in terms of both their exposure and density. It also provided first constraints on the $s$-process astrophysical conditions, for example, in terms of the temperatures that characterises its stellar sites (e.g., \cite{kaeppeler93}). The meteoritic abundances used in those early works characterise the bulk composition of the Solar System\footnote{Note that the bulk $isotopic$ composition of meteorites and planetary objects mostly reflects the average abundances carried by the gas and dust present in the Milky Way interstellar medium at the place and time of the formation of the Sun. The $elemental$ abundances instead are also strongly affected by secondary chemical processing.}. They were reproduced initially by parametric models \cite{kaeppeler89}, and later by stellar models, which were taken to be ``typical'' of the $s$-process sources in the Galaxy \cite{raiteri91,straniero95,gallino98,arlandini99,the07,pignatari10,bisterzo10PaperI}. Today, chemical evolution models of the Galaxy (e.g., \cite{travaglio04,prantzos18,kobayashi20}) are employed to interpret the composition of the Solar System. These models use as input the ejecta of many $s$-process stellar sources, both massive stars and asymptotic giant branch (AGB) stars, of different masses and metallicities, to calculate the evolution of chemical abundances as the Galaxy ages.

On top of this well-mixed, averaged isotopic composition, small nucleosynthetic isotope variations between the average of bulk meteorite groups have also been observed. This indicates that large spatial isotopic heterogeneities were preserved in the early Solar System despite the strong homogenisation of matter that must have occurred during accretion and in the proto-planetary disk. For reviews of this complex and far-reaching topic we direct the reader to \cite{dauphas16,yokoyama16,qin16,bermingham20,kleine20,mezger20}. 
In brief, such heterogeneities are extremely small, with measured variability of the order of $10^{-4} - 10^{-5}$, but they can be measured with sufficiently high precision, $\sim 10^{-6}$, that they can be interpreted as significant signals. This improvement in analytical precision is primarily due to methodological advancements in high-precision multi-collector inductively coupled mass spectrometry (MC-ICPMS), and in the chemical treatment of samples. 
The discovered nucleosynthetic isotope variations could have been inherited from an originally heterogeneous molecular cloud (including roughly 1\% pre-solar dust) and/or represent the product of 
thermal, gravitational, and/or chemical processing in the proto-planetary disk itself (e.g., \cite{trinquier09,steele12,ek20,akram15,hutchison22}). 
Newly synthesised stellar material could also have been added (e.g., \cite{nanne19,lichtenberg21}) during the accretion and/or onto the proto-planetary disk once it had already formed, and the accreting material could have varied with time and/or space according to the composition of the ejecta of potentially nearby stellar objects, e.g., \cite{haba21}.

Possible carriers of the observed nucleosynthetic isotopic variations are ``stardust'' grains, i.e., micro-minerals that formed directly in the ejecta of stars and supernovae and were present in the pre-solar dust inventory of the original solar molecular cloud (which included both stardust and dust formed in the interstellar medium).
Some of these grains are relatively resistant to destruction 
and survived as individual dust grains in the interstellar medium and in the proto-planetary disk. They were preserved inside primitive meteorites that never experienced complete melting, and have been identified and recovered because of their extreme isotopic anomalies \cite{zinner14}. Stardust grains are effectively tiny samples of stellar material and their compositions can be directly compared to stellar model predictions (e.g., \cite{zinner14,nittler16,lugaro18grains,liu19Mo}).
For example, the vast majority of stardust silicon carbide (SiC) grains recovered from meteorites formed in the external layers of C-rich asymptotic giant branch (AGB) stars and carry the signature of the $s$ process in the isotopes of many of the elements heavier that iron, from Sr, Kr, and Zr up to Ba, Nd, and Sm, see, e.g., \cite{gallino97}. In brief, AGB stars are the final phases of the life of low-mass stars (with an initial mass roughly lower than 10 times the mass of the Sun, see review by Karakas \& Lattanzio \cite{karakas14dawes}). They experience H and He burning in their deep layers and the material processed by nuclear reactions is carried to the stellar surface via recurrent mixing episodes. In particular,  He-burning is only partially activated at the bottom of the convective thermal pulses in AGB stars,
producing more carbon than oxygen. Therefore, mixing of this material to the stellar surface allows some of these stars to become C-rich (C$>$O) and produce C-rich dust such as SiC.

\begin{figure}
\centering
  \includegraphics[width=10 cm]{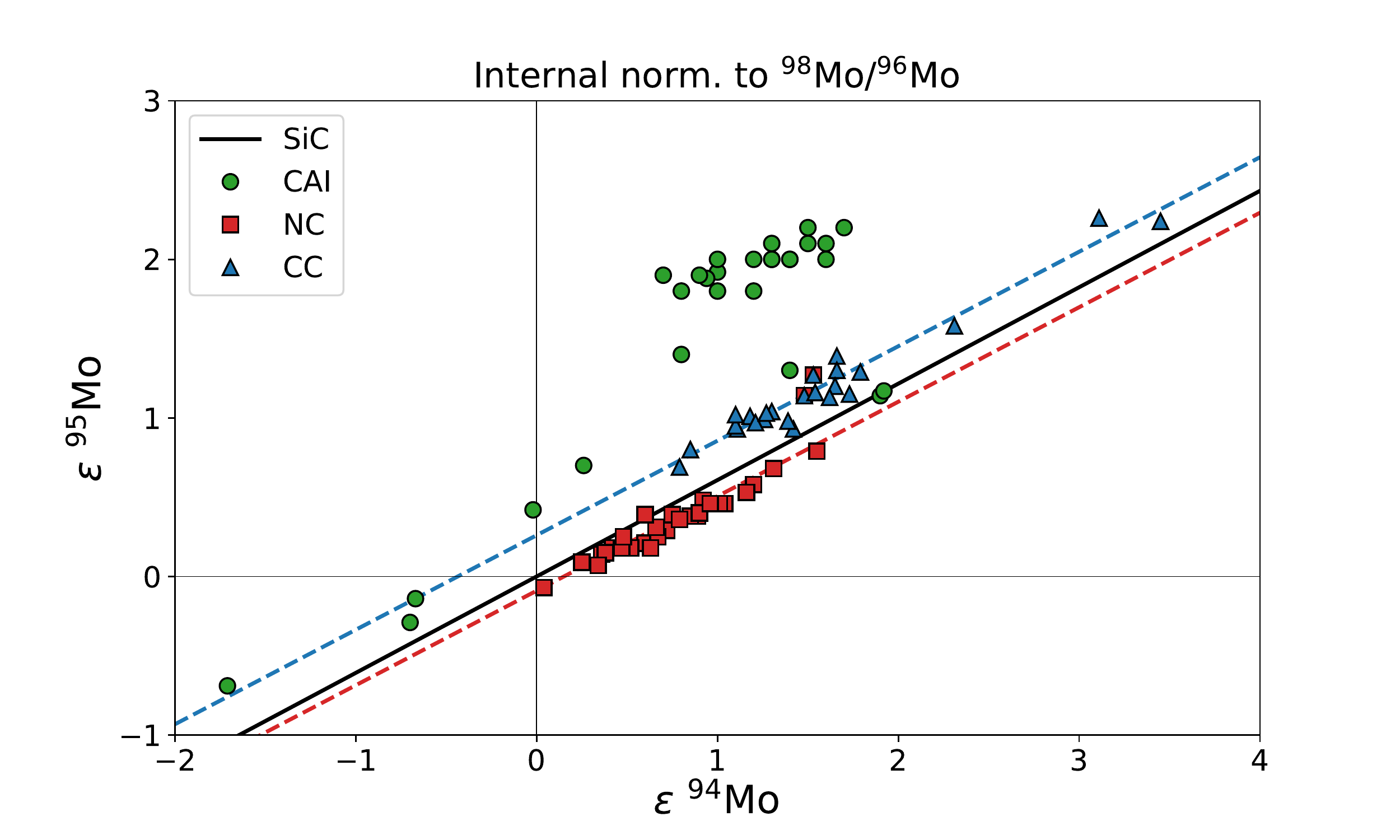}
\caption{Mo isotopic variations measured in different meteoritic objects, shown in the usual representation where isotopic ratios are plotted against each other. The derivation of the $\epsilon$ quantities reported on each axis is described in detail in Section~\ref{sec:internal}. In brief, they are internally normalised isotopic ratios (using \iso{98}Mo/\iso{96}Mo) relative to the laboratory standard used during the measurement (the [0,0] point by definition, close to the natural terrestrial value, but not necessarily identical). 
Error bars are not reported for sake of clarity, with 2$\sigma$ uncertainty for most data points of the order of $\pm$0.1 to $\pm$0.5. The data are from: (i) the compilation of \cite{burkhardt21} for meteoritic bulk rocks classified in the two NC and CC groups according to their composition with respect to observed isotopic dichotomy, with the NC and CC lines from \cite{budde19} ; (ii) the compilation of \cite{brennecka20} for calcium-aluminum-rich inclusions (CAIs), which have sizes of roughly a few cm, are found mostly in primitive meteorites, represent the first solids to have formed in the Solar System, and show the largest deviation from the bulk solar composition; and (iii) \cite{stephan19} for the the best fit trend line for meteoritic mainstream (MS) stardust silicon carbide (SiC) grains that originated from AGB stars, the main sources of the $s$ process in the Galaxy. 
}
\label{fig:Mo}       
\end{figure}

Dauphas et al. \cite{dauphas04} found variations in Mo and Ru in meteorites that clearly follow the pattern of the $s$-process nucleosynthesis that occurs in AGB stars (Figure~\ref{fig:Mo}). Since then, bulk meteorite data has been collected for many more elements (see, e.g., \cite{akram15} and Table~\ref{tab:normalisations}) and the data shows that the Earth carries the largest $s$-process excess identified in the Solar System, for example, for Mo and Ru. This has been interpreted as evidence that the $s$-process-rich material was present in the innermost part of the disk \cite{mezger20,burkhardt21}. 
The $s$-process signatures may have been generated, for example, by thermal effects in the proto-planetary disk as a function of heliocentric distance and/or other disk features. These effects may have resulted in a distribution of stardust SiC grains from AGB stars different from the distribution of pre-solar dust particles that condensed in the interstellar medium. 
For example, the pre-solar SiC stardust that carried the $s$-process signature may have survived better closer to the Sun than the other pre-solar dust \cite{ek20}. 

While the $s$-process variations appear continuous, other heterogeneities are  discontinuous instead. A compositional ``dichotomy'' is observed between different types of meteorites: the non-carbonaceous (NC) and carbonaceous (CC) meteorites that may have formed in the inner ($<$ 3AU) and outer ($>$ 3AU) parts of the disk, respectively \cite{desch18}. This dichotomy is most apparent in the neutron-rich isotopes of intermediate-mass and iron-group elements (e.g., \cite{trinquier07,trinquier09,leya08}), such as $^{48}$Ca, $^{50}$Ti, and $^{54}$Cr; elements heavier than iron affected by neutron captures, such as Mo (Figure~\ref{fig:Mo}) and Ru (e.g., \cite{burkhardt11}); and other isotopes of explosive nucleosynthesis origin, such as \iso{58}Ni (e.g., \cite{steele12}). The dichotomy has been attributed, for example, to the presence of two dust reservoirs in the proto-planetary disk. Although the origin of these variations is unclear, the carriers may have included stardust Cr-rich oxide grains \cite{nittler:18} originating in core-collapse or other types of supernovae \cite{jones:19a,denhartogh22}. Among the possible mechanisms suggested for creating the two reservoirs are, for example, the formation of Jupiter \cite{kruijer17} and the migration of the snow line in the proto-planetary disk \cite{lichtenberg21}. 
While all such speculations seem reasonable, and despite the extensive observational data available, there is no consensus yet on a full scenario that can explain the origin of the large length-scale nucleosynthetic variation along the proto-planetary disk recorded in meteorites.

Molybdenum is an example of an element for which the measured bulk isotope composition is available for a wide range of meteorites, showing both the composition dichotomy between the CC and NC groups and the $s$-process trend (Figure~\ref{fig:Mo}).
We do not discuss the nucleosynthetic origin of the Mo dichotomy here, but note that it is unlikely to originate from the $rapid$ neutron-capture ($r$) process. This is because $r$-process sites in the late Galaxy (such as compact mergers, or special types of supernovae, see, e.g., the review by \cite{cowan21}) are either extremely rare or do not produce the dust needed to preserve and carry an $r$-process anomaly into the Solar System. The nucleosynthetic origin of the Mo dichotomy may be instead related to neutron-capture nucleosynthesis in core-collapse supernovae. These are more common stellar objects well known to produce dust, as also evidenced by the presence of such grains in the stardust inventory, e.g.. \cite{meyer00MoZr,pignatari18,liu18SN}). In general, cosmochemists focus on matching the observed variations using the $s$-, $r$, and $p$-process isotopic abundances predicted for the bulk Solar System \cite{arlandini99,bisterzo11PaperII}. However, the observed variations could also derive from sources that are not major contributors to the Solar System compositions. In fact, Stephan \& Davis \cite{stephan21} proposed that distinct $s$ processes from different AGB star sources may be responsible for the Mo dichotomy.
Figure~\ref{fig:Mo} shows that both the NC and CC groups follow continuous trends on two lines almost parallel to the trend of mainstream (MS) stardust silicon carbide (SiC) grains
(and consistent with an $s$-process signature, as we will show in Section~\ref{sec:Mo}). This supports the idea that
the nucleosynthetic variability within the CC and NC groups may have been controlled by the distribution of SiC grains in the proto-planetary disk \cite{ek20}.

To obtain a full picture, it is necessary to compare models of the $s$ process in AGB stars to meteoritic data for all the available elements, i.e., to add model predictions onto Figure~\ref{fig:Mo}. As presented in detail in the following methodology section, however, this data-model comparison is far from trivial and special considerations are required. In brief, models predictions are given as abundances of each given isotope, while meteoritic data are given as ratios between two isotopes, normalised to another given ratio with an assumed value.  This is usually referred to as ``internal normalisation'' and it is necessary to correct for the ubiquitous presence of mass-dependent fractionation, which does not reflect nucleosynthethic effects. While such transformation involves losing some information, also because  mass-dependent effects may mimic nucleosynthetic effects, it is a necessary step as we demonstrated below.
Furthermore, the meteoritic data do not represent pure stellar nucleosynthetic compositions but only very small deviations towards it and away from the solar composition. This indicates that the stellar material was significantly diluted, which points to a relatively small amount of stardust carriers inside the meteorite relative to the rest of the isotopically solar material.


A straightforward way to compare stellar models with meteorite data is to apply the internal normalisation procedure to the stellar models. This makes it possible to evaluate several stellar models at once in relation to the same meteorite data. There are two approaches to applying the internal normalisation procedure to stellar models. The first approach is to construct a synthetic sample by adding a small amount of stellar abundance to the solar abundance, for any given isotope:

\begin{equation}
    C_{SMP} = C_\odot +  \left(C_{SM} \times x\right)
    \label{eq:intro1}
\end{equation}

\noindent where $C_{SMP}$ is the abundance of the synthetic sample, $C_\odot$ is the solar abundance, $C_{SM}$ is the stellar abundance and $x$ is the dilution factor of the stellar  abundances relative to the solar abundances. The value of $x$ is such that the magnitude of the internally normalised values of $C_{SMP}$ are comparable to that of bulk meteorites, e.g., Simon et al. \cite{simon09} and Figure~\ref{fig:MoExample}. The resulting composition can then be internally normalised. The second approach is to use a linearised version of the internal normalisation procedure that can be applied directly to $C_{SM}$, e.g., Dauphas et al. \cite{dauphas04}. As we show in the Supplementary Materials these two approaches result in almost identical slopes, when close to the solar composition, of the mixing lines between the solar composition and the stellar end-members. Such slopes are the quantity that allows us to compare the models to the data, as we will describe in details in Section~\ref{sec:mixing}. Another way to compare meteorite data with stellar models is to proceed in the reverse way and renormalise the meteorite data to fit the stellar end-members, e.g., Stephan \& Davis \cite{stephan21}. This approach is most useful if the stellar components involved are well constrained, as for Mo where abundant stardust data provide good constraints on the composition of the stellar end-members \cite{stephan21}. We will not discuss this alternative method in further detail here because we aim to provide guidelines for the stellar modellers and nuclear astrophysicists to manipulate their model predictions in order to access the comparison to the meteorite data. 

The aim of this paper is to describe and explain the methodology to compare data and models and highlight possible obstacles and limitations. We will show the results of the internal normalisation procedure as applied to examples of $s$-process abundances to demonstrate that it is not intuitive to make even a qualitative comparison using raw stellar model abundance predictions. In Section~\ref{sec:methodology} we outline the problem and its solutions, in Section~\ref{sec:results} we show some examples of the effect of the procedure described in Section~\ref{sec:methodology} on $s$-process abundances, including the models predictions for Figure~\ref{fig:Mo}. We will summarize our methods and results, draw conclusions, and propose future work in Section~\ref{sec:conclusions}.


\section{Methodology}
\label{sec:methodology}

The determination of stellar signatures hidden within the isotopic composition of meteorites via mass spectrometry is a complex task because of the analytical challenges and because of the presence of effects other than stellar nucleosynthesis that can affect the isotopic abundances. 
High precision isotope analysis generally requires careful chemical separation of the target element because isobaric interferences\footnote{These occur when isotopes of different elements with nearly identical atomic mass are collected together with the target isotopes. Resonance ionization mass spectrometry uses highly specific laser beams and is capable to ionize the target element without isobaric interference. This method have been used to analyse single stardust grains, e.g., \cite{savina03,liu15SrBa}, and chondrules from meteorites \cite{trappitsch18Fe60}.} must be eliminated for accurate results, and because the presence of matrix material can affect both the accuracy and precision of the instrumental analysis \cite{rehkamper12}. There are two types of isotopic variation found in meteorites: mass dependent and mass independent and they are briefly described below. For a more comprehensive discussion on the different types of isotope variations found in meteorites and their components we refer the reader to the reviews by Rehk\"amper et al. \cite{rehkamper12} and Dauphas \& Schauble \cite{dauphas16}.

Mass-dependent isotope fractionation (MDF) results in variations of the order of up to few \%
and arises from physical, chemical, and geological processes that may occur in nature and/or during chemical separation and isotope analysis. As the name indicates, this type of fractionation is a function of the isotope mass and occurs during kinetic fractionation, where light isotopes are often enriched in the reaction product, and equilibrium reactions, where the heavy isotopes are generally enriched in the material with the stiffest chemical bonds \cite{schauble04}. Mass-dependent fractionation occurs during instrumental analysis (see \cite{dauphas16,rehkamper12} for reviews) and chemical separation of the element via ion exchange chromatography, although this can generally be avoided if more than 95\% of the target element is recovered during the chemical procedure. Correcting for MDF is therefore a vital step in the processing of isotope data. This correction, and how it may obscure other isotopic signatures, is explained in the next section. Note that for stardust the measured isotope ratios can be corrected for instrumental MDF using external standards and the remaining natural MDF is insignificant relative to the much larger observed stellar nucleosynthesis anomalies. This  means that, like stellar predictions, stardust compositions cannot be directly compared against high precision data that has been corrected for MDF.

Mass-independent isotope variations are, for example, the stellar nucleosynthetic signatures in bulk meteorites, of the order of $\leq 10^{-4}$, now resolved at the order of $10^{-5}$ to $10^{-6}$. Other examples include effects from the radioactive decay of unstable isotopes, photochemical processes, such as the self-shielding during CO photodissociation invoked to explain O isotope variations in the Solar System,
and interaction of material with galactic cosmic rays, which leads to the production of isotopes via spallation and/or secondary neutron capture reactions \cite{leya13}. The last effect can be corrected using cosmic neutron dosimeters, i.e., isotopes for which such production is well characterised (for example, the isotopes of Pt \cite{hunt17,ek20}). The different size and shape of isotopes can also give rise to nuclear field shift effects \cite{bigeleisen96}, although these typically do not have a significant impact.

\subsection{Internal normalisation procedure}
\label{sec:internal}

When comparing stellar model predictions and stardust compositions to bulk meteoritic data, it is imperative to follow the same data reduction procedure that was used for the meteorite isotopic data set considered and is usually specific to the instrument used to collect the data. In-situ measurements, such as secondary-ion mass spectrometry SIMS routinely employed for stardust, are not able to resolve nucleosynthetic differences between average bulk meteorite groups due to their relatively lower precision. There are two other major types of instrumentation for high-precision analysis of bulk meteorites: thermal-ionization mass spectrometry (TIMS) and multi-collector inductively-coupled-plasma mass spectrometry (MC-ICPMS).  These are capable of producing isotope data of an element with a precision down to a few part per million (ppm) in a $\sim$1 to 100 mg of sample material dissolved in acidic media. The MDF effects from such kind of measurements are usually in the permil range for TIMS and percent range for MC-ICPMS. The exact degree of MDF needs to be determined for each measurement (see, e.g., \cite{rehkamper12,schonbachler16a,schonbachler16b}) and changes with time during analysis. 
The TIMS and MC-ICPMS analyses yield degrees of MDF that are often significantly larger than any observed natural mass-dependent and mass-independent variation (although in some instances nucleosynthetic effects in stardust and leachate samples can be larger), and are routinely corrected for by internal normalisation schemes.

\begin{figure}
  \includegraphics[width=12 cm]{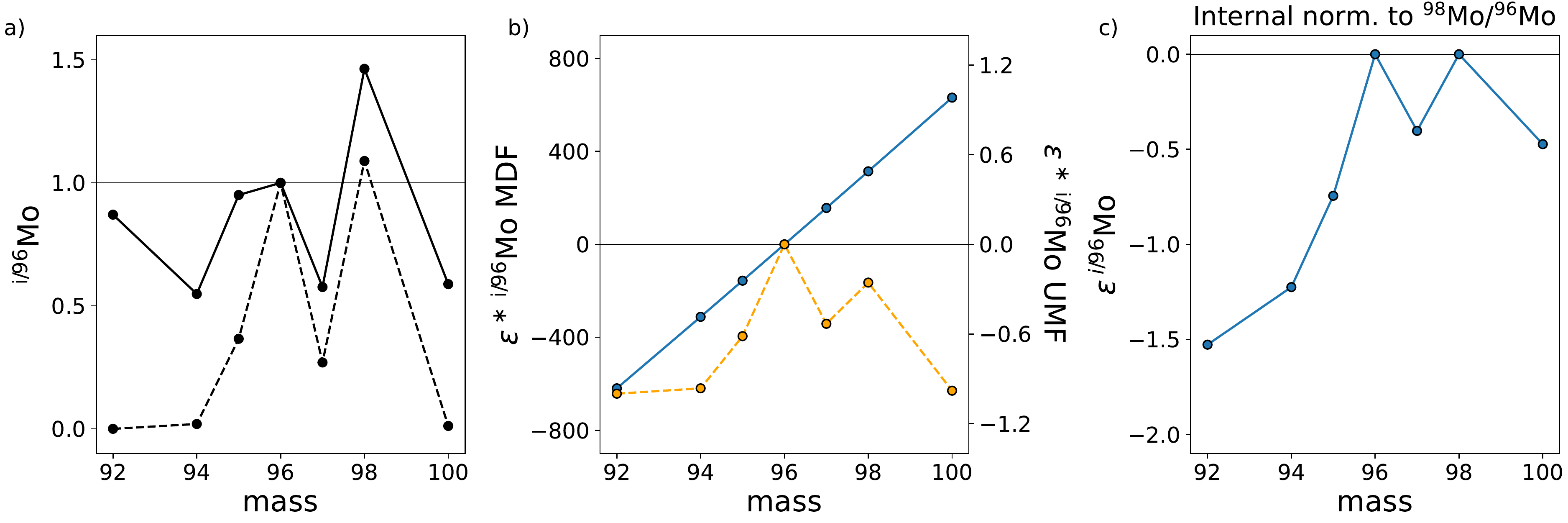}
\caption{Schematic description of the effect of the procedure of internal normalisation on an example $s$-process sample. In the left panel (a), the solid line represents the ratios of the Mo isotopes relative to \iso{96}Mo using the solar abundances, the dashed line represents the same ratios calculated using the $s$-process abundances from mainstream SiC grains \cite{stephan19}. In the middle panel (b), the dashed, orange line (with corresponding values on the right y-axis) represents the $\epsilon$ values of un-mass-fractionated (UMF) $s$-process sample obtained by adding $10^{-5}$ of the $s$-process component to the solar abundances. The solid, blue line (with corresponding values on the left y-axis) represents the $\epsilon$ values of the sample of the dashed line, but with some added mass-dependent fractionation (MDF) obtained by adding a typical MDF effect with $\beta = 1.5$. Note the huge difference between the scale the left and right y-axes. The right panel (c) shows the internally normalized abundance pattern for both the MDF and UMF samples in the middle panel. Note that both lines are identical, which highlights the generality of the internal normalisation method to remove the problem of MDF.}
\label{fig:MoExample}       
\end{figure}

The correction of MDF by internal normalisation assumes that one of the measured isotope ratios is identical to the known/published value of a chosen terrestrial standard. The difference between the measured and the reference value for this particular ratio is then used in the MDF correction procedure. Any mass-independent variation on the normalisation ratio will result in a modification of the MDF correction, and this in turn can drastically change the signature of the resulting mass-independent data, as we will demonstrate in Figure~\ref{fig:MoExample}. The exponential law for isotope fractionation described by Russell et al. \cite{russell78} for Ca is commonly applied to isotope studies of different elements for both TIMS and MC-ICPMS analysis\footnote{Note that, in general, correction of mass-dependent data needs to be carefully made using the appropriate fractionation law. For example, material that has experienced equilibrium effects would not follow the exponential law, and assuming such law in this case would create an artificial mass-independent effect. In many cases, we do not know which law is the correct one to use. Nevertheless, the exponential law has been found to adequately account for the MDF in the majority of studies.}. This laws states that the MDF correction applied to the mass-fractionated measured ratio ($r$)\footnote{Note that in Section~\ref{sec:results}, we will also use the symbol $r$ in Eq.~\ref{eq:method1} for stellar model abundances - even though these ratios do not contain a mass-dependent component.} of two isotopes $i$ and $j$ is derived from the corresponding not mass-fractionated ratios ($R$) as an exponential function of the ratio of their masses $m_i$ and $m_j$:

\begin{equation}
    R_{ij} = r_{ij}{\left(\frac{m_i}{m_j}\right)}^{-\beta},
    \label{eq:method1}
\end{equation}

\noindent where $\beta$ is a measure of the MDF. For TIMS and MC-ICPMS analyses, if $m_i > m_j$, the value of $\beta$ will generally be positive as these instruments favour 
increasing the abundance of the heavier relative to the lighter isotopes\footnote{Eq.~\ref{eq:method1} can also be written as $r_{ij}=R_{ij}(m_i/m_j)^{\beta}$. Sometimes in literature Eq.~\ref{eq:method1} is written as $R_{ij}=r_{ij}(m_i/m_j)^{\beta}$ instead, in which case the value of $\beta$ changes sign relative to Eq.~\ref{eq:method1}, i.e., a positive $\beta$ value in Eq.~\ref{eq:method1} would be a negative $\beta$ in this variation.}.




Since isotopes of the same element have identical chemical properties, all isotopes during a single measurement follow the same mass fractionation law, i.e., the value of $\beta$ is the same. The $j$ reference isotope at denominator is usually chosen such that it has a high abundance and no isobaric interferences, in order to minimize statistical uncertainties and the error correlation between $r_{ij}$ and $r_{kj}$, the ratio used for internal normalisation.  If we then consider Eq.~\ref{eq:method1} but applied to the ratio $kj$ used for internal normalisation, where the denominator isotope is by convention always the same, we can extract the exponent ($-\beta$) , equate the two expressions, and calculate $R_{ij}$ of a sample (SMP) as:

\begin{equation}
    R^{\rm{SMP}}_{ij} = r^{\rm{SMP}}_{ij}{\left(
    \frac{R^{\rm{STD}}_{kj}}{r^{\rm{SMP}}_{kj}}
    \right)}^
    \frac{\ln{(m_i)} - \ln{(m_j)}}{\ln{(m_k)} - ln{(m_j)}},
    \label{eq:method2}
\end{equation}

or, changing the final step of the derivation $e^{ab}=(e^a)^b$ to have the exponent $=-\beta$, as:

\begin{equation}
    R^{\rm{SMP}}_{ij} = r^{\rm{SMP}}_{ij}{\left(
    \frac{m_i}{m_j}
    \right)}^
    \frac{\ln{(R^{\rm{STD}}_{kj})} - \ln{(r^{\rm{SMP}}_{kj})}}{\ln{(m_k)} - ln{(m_j)}},
    \label{eq:method2b}
\end{equation}

\noindent  
where it is assumed that $R^{\rm SMP}_{kj} = R^{\rm STD}_{kj}$, i.e., 
the reference value used for the internal normalisation, usually the terrestrial standard (STD). 
Note that an element needs at least 3 stable, measurable, isotopes to account for the MDF using this approach. If the final aim is to derive the nucleosynthetic effect, then the normalising isotopes should also be free of other mass-independent effects, such as radiogenic decay (see, e.g., discussion in Sec.~\ref{sec:Sr}). 

Measured deviations from the terrestrial standard ratios are usually so small that they are not given as percent, i.e., deviation in parts per 100, but as deviation in parts per 1000, also called permil $\delta$ (and typical for stardust data), as parts per ten thousands, called $\epsilon$, or parts per million, called $\mu$. For example:

\begin{equation}
    \epsilon R^{\rm{SMP}}_{ij} = {\left(
    \frac{R^{\rm{SMP}}_{ij}}{R^{\rm{STD}}_{ij}} - 1
    \right)} \times 10^4.
    \label{eq:method3}
\end{equation}

\noindent The ratio used for internal normalisation can be indicated in different ways, for example, by using the last digit of the mass of each isotope in the ratio within brackets, sometimes separated by a comma. It is also frequently omitted from the notation if all data presented uses the same internal normalisation ratio. In the figures presented here we explicitly indicate the normalising ratio at the top of each plot or in the figure caption (e.g., Fig.~\ref{fig:Mo}).
Combining equations \ref{eq:method2} and \ref{eq:method3}, and remembering that $(a/b)^c$ is the same as $(b/a)^{-c}$, gives:

\begin{equation}
    \epsilon R^{\rm{SMP}}_{ij} = {\left[
    {\left(
    \frac{r^{\rm{SMP}}_{ij}}{R^{\rm{STD}}_{ij}}
    \right)}
    {\left(
    \frac{r^{\rm{SMP}}_{kj}}{R^{\rm{STD}}_{kj}}
    \right)}^{-Q_i} - 1
    \right]} \times 10^4,
    \label{eq:method4}
\end{equation}

\noindent where $Q_i$ is the exponent as given in Eq.~\ref{eq:method2}. We further define $\epsilon^{\ast} r_{ij}$ as the composition of a sample that has not been corrected for MDF (i.e., calculated using $r_{ij}$ instead of $R_{ij}$ in Eq.~\ref{eq:method3}).

In Figure \ref{fig:MoExample} we use Eq.~\ref{eq:method4} to illustrate the effect of internal normalisation for the example of an $s$-process sample with an added small mass-independent signal. In the left panel (a) we show the solar abundance  \cite{lodders09} (solid line) and the $s$-process component from mainstream SiC grains \cite{stephan19} (dashed line). 
In the middle panel (b) we show a sample created by adding $10^{-5}$ of the $s$-process component to the solar composition to recreate a small $s$-process excess (dashed, orange line). Then, we mass-fractionate the same sample to a degree typical of that seen for MC-ICPMS instruments ($\beta = 1.5$), to create a synthetic meteorite isotope data point (solid, blue line). The variation due to mass-fractionation is two to three orders of magnitude larger than the nucleosynthetic variation. This highlights the need of the correction for mass-dependent fractionation to ascertain the nucleosynthetic composition of the sample, which is often orders of magnitude smaller. Finally, in the right panel (c) of Figure \ref{fig:MoExample} we show the sample after it has been internally normalised to the solar abundance of \iso{98}Mo/\iso{96}Mo. The small negative offset caused by the $s$-process contribution on \iso{98}Mo/\iso{96}Mo (dashed line in the middle panel of Figure \ref{fig:MoExample}) results in a $\beta$ value calculated from the sample of $1.4988$, instead of $1.5$. This small difference represents the effect of the nucleosynthetic component and results in a small under-correction of the mass-dependent component. Note that the size of this under-correction is independent of the actual mass fractionation experienced by the sample. This is because the apparent $\beta$ value of the sample is always equal to the true $\beta$ ($1.5$ in this example) plus the difference due to the mass-independent component ($-0.0012$ in this example). Therefore, after internal normalisation (Figure \ref{fig:MoExample}c) we obtain the same abundances starting from both lines in panel b, i.e., we obtain the same abundances in both cases, if mass fractionation was added or not. The final result of the internal normalisation procedure is that the $s$-process pattern is rotated anti-clockwise around the normalising isotope (Figure \ref{fig:MoExample}c). 

The larger the mass difference between each isotope and the normalising isotope is, the larger the rotation effect will be,  because the applied correction factor scales with the mass difference between the numerator and denominator isotopes. While the internally normalised $s$-process composition is different from the ``true'' $s$-process pattern, it provides model predictions in a representation directly comparable to the meteorite data reported in literature. 

Table \ref{tab:normalisations} reports a list of elements heavier than and including Fe, for which nucleosynthetic variations are identified in bulk meteorites, together with the ratios used for the internal normalisation and some recent references. The normalising ratio is usually chosen to optimise the analytical resolution of the small isotope anomalies present in meteorites (see, e.g., the discussion in \cite{burkhardt11}). The normalising ratio can also be chosen to minimize, or to enhance, the rotational effect resulting from an expected mass-independent signature. For example, the \iso{98}Mo/\iso{96}Mo ratio commonly used for Mo data minimises the rotational effect due to $s$-process variations because both isotopes are significantly produced by the $s$ process. For elements for which stardust SiC data is available, the normalising isotope used and example references are also reported in Table \ref{tab:normalisations}. 

\begin{table}
\caption{Iron and heavier elements that show nucleosynthetic heterogeneity in bulk meteorites and CAIs, the isotopic ratios most commonly used for the internal normalisation, and some example references. When also stardust SiC grains data is available, the isotope at denominator is reported together with example references.}
\label{tab:normalisations}
\begin{tabular}{cccccc}
\hline\noalign{\smallskip}
 & \multicolumn{3}{c}{Internal Normalisation} &  \multicolumn{2}{c}{Stardust} \\
 Element   & Ratio   &    Value & Reference   &   Isotope & Reference   \\
\noalign{\smallskip}\hline\noalign{\smallskip}
 Fe        & 57/54    & 0.362549 & \cite{hopp22}                & 56  & \cite{trappitsch18}   \\
           & 57/56    & 0.023095 & \cite{hopp22}                & & \\
 Ni        & 61/58    & 0.016744 & \cite{cook21}                & 58  &  \cite{trappitsch18}    \\
           & 62/58    & 0.053389 & \cite{cook21}                & & \\
           & 62/61    & 3.188500 & \cite{steele12} & & \\
 Zn        & 67/64    & 0.082160 & \cite{steller22}             & & \\          
           & 68/64    & 0.385564 & \cite{savage22}              & & \\
 Sr     & 86/88    & 0.119400 & \cite{charlier19}            & 86 & \cite{liu15SrBa} \\
 Zr        & 94/90    & 0.338100 & \cite{akram15}               & 94  & \cite{barzyk07,nicolussi97}    \\
 Mo          & 98/96    & 1.453173 & \cite{spitzer20}             &  96 &  \cite{nicolussi98}    \\
 Ru        & 99/101   & 0.745075 & \cite{fischer17}             & 100 &  \cite{savina04}   \\
 Pd        & 108/105  & 1.188990 & \cite{ek20}                  & &    \\
 Ba        & 134/136  & 0.307800 & \cite{andreasen07}           & 136 & \cite{liu14Ba}   \\
           & 134/138  & 0.033710 & \cite{bermingham16}          & &    \\
 Nd        & 146/144  & 0.721900 & \cite{frossard21}             & 144 & \cite{yin06} \\
 Sm          & 147/152  & 0.560830 & \cite{burkhardt15Nd}           & 149 &  \cite{yin06}  \\
 Er        & 166/168  & 1.241400 & \cite{shollenberger18}       & 168 & \cite{yin06}    \\
 Yb        & 174/172  & 1.477200 & \cite{shollenberger18}       & 172 & \cite{yin06}    \\
 Hf        & 179/177  & 0.732500 & \cite{shollenberger18}       & 178 & \cite{yin06}    \\
           &          &          &                              & 180 & \cite{avila12}   \\
 W         & 186/183  & 1.985900 & \cite{kruijer17}             & 184 & \cite{avila12}    \\
           & 186/184  & 0.927670 & \cite{kruijer17}             & &    \\
\noalign{\smallskip}\hline
\end{tabular}
\end{table}

\subsection{Calculating mixing lines}
\label{sec:mixing}

\begin{figure}
\centering
  \includegraphics[width=7 cm]{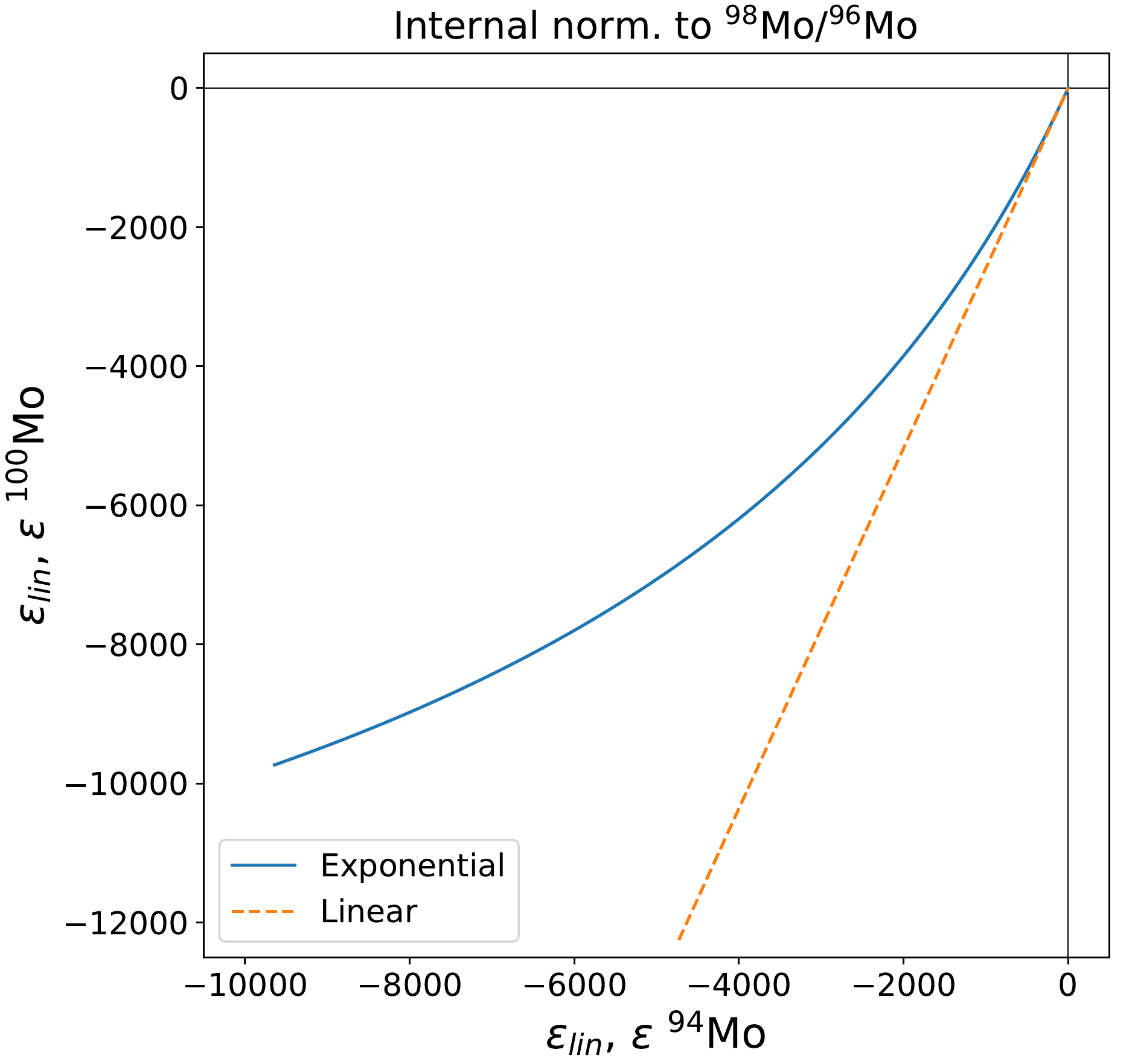}
\caption{Mixing lines calculated for Mo between the solar composition and the $s$-process composition, represented by mainstream SiC grains \cite{stephan19}, using the exponential (Eq.~\ref{eq:method4}; Blue solid line) and linear (Eq.~\ref{eq:method5}; Orange dashed line) equations. The slopes of the two mixing lines overlap only when the offset from the solar composition (point [0, 0]) is of the order of a few 100 $\epsilon$ or less. Bulk meteorites exhibit anomalies well within this range, therefore the linear method is a valid approximation to compare the models to these data.}
\label{fig:expvslin}       
\end{figure}

There are stark differences between meteorite data and stellar models due to the fact that the meteorite data do not represent pure stellar compositions (as in the case of stardust grains), but material in which the stellar signature was strongly diluted with material carrying the standard solar composition. In fact, the stellar composition is usually located far outside the boundaries of a plot as that shown in Figure \ref{fig:Mo} and needs to be further converted into the slope of a line that passes through the point representing material of the standard, terrestrial, or solar composition (usually, but not necessarily, the zero point). 
These lines are called ``mixing lines'' and represent mixtures between the solar, or terrestrial, isotope composition and that of the considered stellar source.
Points along these lines represent compositions with different dilution of the stellar source material, with the stronger the solar dilution, the closer the point to the zero point.
When considering isotopes of the same element, mixing lines are straight in ``three-isotope plots'', i.e., two isotopic ratios with the same denominator plotted against each other. This is the case, for example, for the comparison between not internally normalised stardust and stellar model compositions. However, it is not necessarily the case for mixing lines between the extremely different internally normalised compositions considered here (stellar and solar). In this case the mixing line is curved because the magnitude of the internal normalisation correction scales exponentially, and not linearly, relative to the difference between the true ratio and the assumed ratio.
Therefore, in this case the slope computed close to the point representing the internally normalised stellar abundances can be significantly different from the slope close to the [0,0] point (Figure \ref{fig:expvslin}).

To correct the stellar model predictions for MDF and finally derive the slope of the mixing line that can be compared to that derived from meteorite data we can reasonably approximate $\epsilon$ from  Eq.~\ref{eq:method4} to the following linear $\epsilon_{\rm lin}$ formula\footnote{This can be done for example by writing ($r^{\rm{SMP}}_{ij}/R^{\rm{STD}}_{ij}$) as ($1+10^{-4} \epsilon^{\ast} r^{\rm{SMP}}_{ij}$) from Eq.~\ref{eq:method4}, and equivalently for the $kj$ ratios, where the asterisk indicates that the value is calculated from $r^{\rm{SMP}}_{ij}$ rather than $R^{\rm{SMP}}_{ij}$ as in Eq.~\ref{eq:method3}, and then using the first two terms of the Taylor expansion $(1+x)^{-Q}=(1-Qx)$ followed by some simple algebra and the removal of the second order term multiplied by $10^{-8}$.}:

\begin{equation}
    \epsilon R^{\rm{SMP}}_{ij} \simeq \epsilon_{\rm lin}R^{\rm{SMP}}_{ij} = {\left[
    {\left(
    \frac{r^{\rm{SMP}}_{ij}}{R^{\rm{STD}}_{ij}} -1
    \right)} - Q_i
    {\left(
    \frac{r^{\rm{SMP}}_{kj}}{R^{\rm{STD}}_{kj}} - 1
    \right)}
    \right]} \times 10^4.
    \label{eq:method5}
\end{equation}

\noindent The slope is then determined by dividing the $\epsilon_{\rm lin}$ values calculated for two different isotopes. Note that masses considered in cosmochemisty are usually not rounded, for example, for \iso{86}Sr the accurate value of 85.9092606 u is used, etc. 
For the purpose of calculating mixing lines for comparison with meteorite data, stellar abundances can be used for the calculation of the slope directly using $\epsilon_{\rm lin}$ from Eq.~\ref{eq:method5}, without the need to construct a synthetic sample as we did in Figure~\ref{fig:MoExample}.

As illustrated in Fig.~\ref{fig:expvslin}, the $\epsilon_{\rm lin}$ values calculated from Eq.~\ref{eq:method5} are by definition linear approximations and should be used only to derive the slope close to the [0,0] point representing the solar composition (as done in Figure~\ref{fig:Srexpl}). 

To derive the slope close to the [0,0] point using the $\epsilon$ values calculated from Eq.~\ref{eq:method4} instead, it is necessary to create a synthetic sample by diluting the stellar abundances with the solar abundance using Eq.~\ref{eq:intro1} (see, e.g., Fig.~\ref{fig:MoExample}, and Simon et al. \cite{simon09}). 

We investigated the difference between $\epsilon$ and $\epsilon_{\rm lin}$ for a synthetic sample created using a selection of stellar models from the FRUITY database \cite{cristallo11FRUITY} in combination with solar compositions from Lodders et al. \cite{lodders09} and Anders \& Grevesse \cite{anders89}. The values calculated for the comparisons of Sec.~\ref{sec:results} are available in the Supplementary Material. We find that there is virtually no difference between mixing lines calculated using $\epsilon$ (Eq.~\ref{eq:method4}) on a synthetic sample (from Eq.~\ref{eq:intro1}), or using $\epsilon_{\rm lin}$ (Eq.~\ref{eq:method5}) directly on the stellar abundances.  Furthermore, the choice of the normalising solar/terrestrial ratios does not have a large effect on the calculated mixing lines because the stellar abundance ratios are typically very different from them. However, some variation between using $\epsilon$ and $\epsilon_{\rm lin}$ are found for large mass differences (i.e., when $Q$ is large) and for models that produce variations of two orders of magnitude or more. Generally, to calculate mixing lines it is sufficient to use the same solar values taken in the model calculations to represent the initial composition of the star, even if they are not identical to the reference values in meteorite studies reported in Table~\ref{tab:normalisations}. Although, one needs to be careful of artifacts that can appear for models where no, or little nucleosynthesis takes place if the solar abundances used in the models as starting compositions differ slightly - even if only in the fifth digit - from those used in the calculation of $\epsilon$ value. This will produce spurious values different from zero and therefore erroneous numbers for the slopes. We also compared $\epsilon_{\rm lin}$ to the linearised equations in Dauphas et al. \cite{dauphas04} frequently used in the literature to calculate mixing lines. The approach presented here and that of Dauphas et al. \cite{dauphas04} are mathematically identical with the exception that Dauphas et al. \cite{dauphas04} further simplifies the calculation of $Q_i$ by omitting the natural logarithm of the masses, thus approximating $Q_i$ as $(m_i-m_j)/(m_k-m_j)$ (Eq. (3) in \cite{dauphas04}). This simplification does not significantly affect the mixing lines for most models. The largest differences occur for the largest mass differences and for the largest differences in magnitude between the smallest and largest variations. We recommend using the natural logarithm of the masses in Eq. (3) of \cite{dauphas04}, which yields identical mixing lines to those calculated using Eq.~\ref{eq:method5}.


A more complex situation arises when isotopes of different elements are plotted against each other because in this case variations due to chemistry can also play a role. 
This is referred to as a four-isotope plot in geochemistry and the resulting mixing lines are hyperbolas. Here the mixing relationship not only depends on the isotopic compositions, but also on the elemental ratio of the elements involved. The more this elemental ratio differs between Solar System material and the stellar material, the stronger is the curvature of the hyperbola. There are two ways that the stellar elemental ratios can differ from solar. First, the nucleosynthesis that affects the isotopic ratios may also affect the elemental ratios. The curvature of the hyperbola can be determined directly from the stellar nucleosynthesis models as the ratio of the over-abundances, relative to solar, of the two isotopes at denominator. Close to zero, i.e., for high dilutions, such hyperbola can still be approximated by a straight line and it is possible to consider this effect by simply multiplying the slope derived from $\epsilon_{\rm lin}$ (using Eq.~\ref{eq:method5}) by the ratio of the over-abundances, relative to solar, of the two isotopes at denominator, as in the case of Mo and Ru discussed by \cite{dauphas04}. Second, the chemistry of dust formation may affect the relative abundance of the different element in a stardust grain. In this case, a further factor needs to be introduced to account for the fact that different elements may be incorporated into stardust differently, also depending on the environment and the type of dust. The difficulty here in defining the effect of chemical processes is that stardust condenses in situations of non-equilibrium chemistry; moreover, for many trace elements we do not know how their condensation works. In these cases, not discussed here further, we need to be guided by observations and reasonable assumptions. For example, in the case of the Cr-rich oxide grains, den Hartogh et al. \cite{denhartogh22} used the Al:Mg ratios measured in those grains \cite{dauphas:10}. For heavy $s$-process elements in SiC some data on elemental abundances are available \cite{amari95b}, for the case of Pd, however, no quantitative data is available, and only a qualitative behaviour was considered by Ek et al. \cite{ek20} to interpret the data.

\subsection{Further considerations on the comparison between data and models}
\label{sec:further}

For comparison to the bulk meteoritic data, local stellar abundances should be used at the time and location where the carrier of the observed anomalies (e.g., stardust) potentially formed. These are different from the total integrated stellar yields (i.e., the abundances integrated over the total mass ejected) required by models of the chemical evolution of the Galaxy. For AGB stars, the final surface abundances are a good approximation to consider for each stellar model (for example, of a given mass and metallicity). This is because there is a high degree of both spatial and temporal homogeneity in the outer regions where dust forms. Spatially, the whole envelope of the star is convective, and therefore material is well mixed. Temporally, most of the mass loss occurs in the very final phases of the evolution, therefore, most of the stardust grains carry such a final composition. Also when considering core-collapse supernova models, the local abundances as function of the location in mass within the ejecta should be used, rather than the total yields or the surface composition. 
This is because dust formation probably occurs before the material in the different mass layers of the ejecta is microscopically mixed with all the layers (see discussion in \cite{pignatari13grains,denhartogh22}). 
Therefore, while for AGB stars, one slope well represents one stellar model, for core-collapse supernovae, it is necessary to plot the slopes of the mixing lines derived from the compositions at different mass coordinate \cite{steele12,hopp22}. Different mass shells of a given supernova experience different nucleosynthesis and variable slopes are produced. Total stellar yields are useful if it is of interest to consider dust formed in the interstellar medium, after all stellar yields are homogenised. In this case, however, it is not single star models that should be considered but their implementation into the global modelling of the chemical evolution of the Galaxy, or at least of a stellar population. This is because the interstellar medium is built by many different stellar sources. 

Most previous work on the $s$-process elements have considered the $s$-, $r$, and $p$-component abundances  presented by Arlandini et al. \cite{arlandini99} and Bisterzo et al. \cite{bisterzo11PaperII}. However, these studies were targeted at matching the bulk $s$-process distribution of the Solar System. The observed anomalies could derive, instead, from specific astrophysical sources that may not follow such distribution. In fact, stardust SiC grains from AGB stars are well known to present $s$-process isotopic features that are not the same as those of the solar distribution \cite{ott90,hoppe97}. For example, matching the grain data requires a lower time-integrated neutron flux - resulting in lower \iso{88}Sr/\iso{86}Sr and \iso{138}Ba/\iso{136}Ba ratios - than that which produced the solar $s$-process abundance pattern (see, e.g., \cite{lugaro18grains,liu18c13pocket,lugaro20} and Table~\ref{tab:models}). Therefore, specific models of AGB stars of different masses and metallicities need to be compared to stardust and bulk meteoritic data because each stardust grain can in principle originate from a different star. The same consideration in principle applies to the anomalies referred in the literature as to $r$- and $p$-process anomalies. The main $r$- and $p$-process sites in the Galaxy may be copious and variable over the history of the Galaxy: for the $r$ process they range from, e.g., neutron stars mergers to magnetars \cite{cowan21}, for the $p$ process they may also include thermonuclear supernovae \cite{travaglio18}. Furthermore, anomalies that mimic the $r$ and the $p$ processes may be also produced in astrophysical sites that are not the major contributors of these isotopes in the Galaxy, but that can still affect relative isotopic abundances and be the site of dust formation. For example, the neutron bursts (the $n$ process) in the He shell of core-collapse supernovae and possible explosive neutrino-wind components produce a peculiar signal that for certain isotopic ratios may look similar to the $r$ process (e.g., \cite{farouqi09,pignatari18,bliss18,kratz19}). 

Finally, we note that for core-collapse supernovae most of the comparisons available in the literature used the models by Rauscher et al. (\cite{rauscher02}, although see an exception in den Hartogh et al. \cite{denhartogh22}). It is well known that core-collapse supernovae models are very uncertain therefore it is unsatisfactory that typically just one set of models has been employed. More models, with their detailed composition as function of mass, needs to be made available for this task. One of the main aims here is to illustrate for the nuclear astrophysics community the significance of stellar model predictions within the framework of meteorite data. Working Package 9 of the ChETEC-INFRA \url{www.chetec-infra.eu} project (funded by the European Union’s Horizon 2020 research and innovation programme) has the aim to provide the tools for the comparison so that more researchers can contribute to the field of cosmochemistry.


\section{Examples}
\label{sec:results}

To illustrate the effect of the internal normalisation on the $s$-process predictions from AGB models we present some examples for two selected elements: Sr and Mo, using the data from the models of \cite{lugaro18grains}. To generate the plots we use the open source code available via ChETEC-INFRA at  \url{https://www.chetec-infra.eu/resources/#ToolsMeteorites}. Among the models presented by \cite{lugaro18grains}, we consider first that of initial mass 3 \msun\ and solar metallicity (Z=0.014 from \cite{asplund09}) as a typical $s$-process result, and then compare it to two more models of the same and of higher mass (3 and 4 \msun) and twice-solar metallicity (Z=0.03). The over-abundances, relative to the initial values, of the isotopes of interest here for these three models are reported in Table~\ref{tab:models}.

\begin{table}
\caption{Over-abundances, i.e., final production factors at the stellar surface relative to the initial values, of the isotopes of interest for the three AGB models considered here. For the Z=0.014 model, the initial values are the solar values given by \cite{asplund09}, for the Z=0.03 models, they are the same values multiplied by roughly 2, by definition, which results in lower over-abundances, but does not impact the isotopic ratios. The $p$-only \iso{84}Sr and \iso{92}Mo, and the $r$-only \iso{100}Mo are not included in the network of \cite{karakas16} and are assumed here to have production factors equal to unity in all the models, which is accurate within a few percent (see also \cite{lugaro18grains} for discussion on \iso{84}Sr). Also shown are their ratios relative to another isotope (in brackets), which is the first step for the calculations of $\epsilon$ values. }
\label{tab:models}       
\begin{tabular}{llll}
\hline\noalign{\smallskip}
Isotope & M=3 \msun; Z=0.014 & M=3 \msun; Z=0.03 & M=4 \msun; Z=0.03 \\
\noalign{\smallskip}\hline\noalign{\smallskip}
\iso{86}Sr (/\iso{88}Sr) & 26 (0.81) & 20 (1.18) & 10 (1.02) \\
\iso{87}Sr (/\iso{86}Sr, /\iso{88}Sr) & 25 (1.25, 0.78) & 19 (0.95, 1.12) & 9.5 (0.85, 0.97) \\
\iso{88}Sr (/\iso{86}Sr) & 32 (1.23) & 17 (0.85) & 9.8 (0.98) \\
\iso{94}Mo (/\iso{96}Mo) & 1.28 (0.03) & 1.18 (0.08) & 0.99 (0.13) \\
\iso{95}Mo (/\iso{96}Mo) & 24 (0.63) & 8.3 (0.59) & 5.5 (0.73) \\
\iso{96}Mo & 38 & 14 & 7.5 \\
\iso{97}Mo (/\iso{96}Mo) & 21 (0.55) & 8.5 (0.61) & 4.2 (0.56) \\
\iso{98}Mo (/\iso{96}Mo) & 30 (0.79) & 11 (0.79) & 5.0 (0.67) \\
\noalign{\smallskip}\hline
\end{tabular}
\end{table}

\subsection{Strontium}
\label{sec:Sr}

Due to mass-independent variations of \iso{87}Sr as the result of the radiogenic decay of the long-lived isotope \iso{87}Rb (T$_{1/2}$=49 Gyr), current bulk meteoritic data for Sr have limited application for nucleosynthetic studies. However, we chose this element as a first example because it has peculiar features from the point of view of the $s$-process nucleosynthesis and only 4 stable isotopes, at masses 84, 86, 87, 88.
This makes Sr a useful element to demonstrate the effect of the internal normalisation on $s$-process abundance because for any given internal normalisation there are only two ratios that can be plotted against each other and therefore only one plot. 

\begin{figure}
\centering
  \includegraphics[width=12 cm]{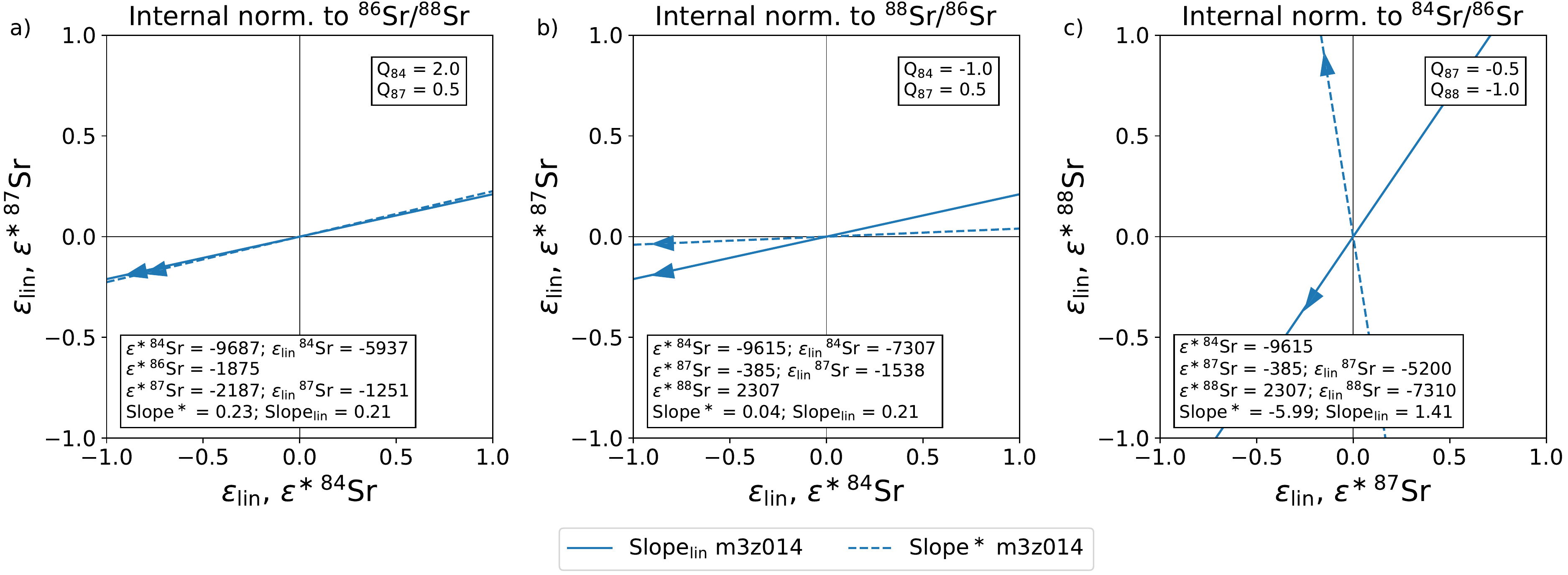}
\caption{Example for calculating the slope of the mixing line in the vicinity of the Solar System end-member (the [0,0] point) using the $s$-process predictions at the surface of an AGB star of initial mass 3 \msun\ and solar metallicity (Z=0.014) at the end of the evolution from Table~\ref{tab:models}. Solid and dashed lines represent, respectively, the slopes derived using $\epsilon_{\rm lin}$ (i.e., the linear approximation Eq.~\ref{eq:method5}) and 
$\epsilon^{\ast}$ (i.e., derived from applying Eq.~\ref{eq:method3} directly to the stellar abundance ratios). Here we considered 3 different possible choices of the normalising ratios, each following the top label ``Internal norm. to''. The arrows on each mixing line indicate the direction of the $s$-process end-member. The numerical values needed to calculate $\epsilon_{\rm lin}^{\ast}$ are also reported, to support the clarity of the description in the text. Slope$_{\rm lin}$ and Slope$^\ast$ in the legend of this figure, and following figures, refers to the slope calculated using $\epsilon_{\rm lin}$ and $\epsilon^\ast$, respectively. As the normalising ratios are indicated separately in each plot, in this and the following figures $\epsilon$ values are followed only by the isotope at numerator.} 
\label{fig:Srexpl}       
\end{figure}

In principle, given the 4 isotopes, Sr offers 12 possibilities of internal normalisation choices, i.e., twice the number of possible combinations of 2 isotopes out of 4. However, it is a general feature that
the slope of the line calculated using the internally normalised ratios remains the same if we consider one normalising ratio or its reverse (e.g., for \iso{86}Sr/\iso{87}Sr or \iso{87}Sr/\iso{86}Sr, as shown in Figure~\ref{fig:Srexpl}). 
 Therefore, generally speaking the number of possible combinations given a number of isotopes $n$ is equal to $n$!/2($n-k$)!, where $k$=2 and $n$ is the number of isotopes ($n$=4 in the case of Sr).

Of the four isotopes, \iso{84}Sr is a pure product of the $p$ process, therefore its abundance at the surface of AGB stars is unchanged, within a few percent, relative to the initial abundance. The other 3 isotopes, instead, are mostly of $s$-process origin, specifically \iso{88}Sr has a magic number of neutrons (50) and represents, with Y and Zr, the first $s$-process peak. \iso{86,87}Sr are produced both in massive stars and in AGB stars (with roughly half and half contributions to their solar abundances), while \iso{88}Sr is most efficiently produced in AGB stars\footnote{A debate about a possible $\sim$20\% contribution from massive stars to the $s$-process first peak elements is ongoing. Such a debate does not crucially affect our discussion here because we compare data to single AGB models rather that to the full chemical evolution of the Galaxy.} \cite{travaglio04,cristallo15LEPP,prantzos20}. 

Strontium is also interesting because the \iso{88}Sr/\iso{86,87}Sr isotopic ratios vary the most among all the $s$-process elements when changing the metallicity of the AGB star \cite{lugaro20}. Models of metallicity higher than solar are a better match to the composition of large ($> 1 \mu$m) SiC grains - especially for Sr and Zr, but also for Ba - and their composition differ from that of their solar metallicity counterpart, as discussed at length previously \cite{lugaro18grains,lugaro20}. This result is also clearly shown in Table~\ref{tab:models} where \iso{88}Sr is more over-produced than the other Sr isotopes at solar metallicity, while it has a similar, or even lower, production factor at the higher metallicity. It is therefore interesting to investigate if these different $s$-process predictions are still resolved by the different mixing lines derived within the representation discussed above, and, if so, under which choices of the internal normalisation these variations are more or less evident.

Figure~\ref{fig:Srexpl} presents three example out of the possibilities of internal normalisation choices mentioned above. The $\epsilon^{\ast}$ and $\epsilon$ values can be easily derived from the values of the over-abundance relative to solar provided in Table~\ref{tab:models} using Eq.~\ref{eq:method3} and Eq.~\ref{eq:method5}, respectively. The transformation between $\epsilon^{\ast}$ (derived without internal normalisation) and $\epsilon_{\rm lin}$ (derived with internal normalisation) clearly shows that the slopes of the mixing lines can be more or less modified depending on all the different factors involved. For example, the subtraction term in Eq.~\ref{eq:method5} can change sign depending on the values of both the relevant $\epsilon^{\ast}$ and $Q$ values. Therefore, if the normalising ratio is \iso{86}Sr/\iso{88}Sr (left panel, a), all the $Q$ values are positive, all the $\epsilon_{\rm lin}$ and $\epsilon^{\ast}$ values remain strongly negative, and the slope does not change significantly. If the normalising ratio is \iso{88}Sr/\iso{86}Sr (middle panel, b), not all the $\epsilon^{\ast}$ are negative and not all the $Q$ are positive, which result in a modification of the slope. The left (a) and middle (b) panels demonstrate that reversing the normalising ratio does change all the values of $\epsilon_{\rm lin}$ and $\epsilon^{\ast}$, however, the final slope from the ratio of the $\epsilon^{\ast}$ values remains unchanged. Note that in these cases the direction of the $s$-process excess is always towards negative values, as indicated by the arrows in the figure, and data points located on the line in the quadrant of positive values would represent material with an $s$-process deficit. The right panel (c) shows a more dramatic effect from using the $p$-only \iso{84}Sr as the internally normalising isotope and the resulting negative $Q$ values. In this case the strongly negative $\epsilon^{\ast}$ used for the normalisation, coupled to the negative $Q$ values, produce an inversion of the sign of the $\epsilon^{\ast}$ value corresponding to \iso{88}Sr. While the examples presented here are far from exhaustive, they clearly demonstrate that the representation of the predicted abundances for comparison to meteoritic data are far from intuitive and need to be accurately calculated for each case before conclusions can be drawn.

\begin{figure}
  \includegraphics[width=12 cm]{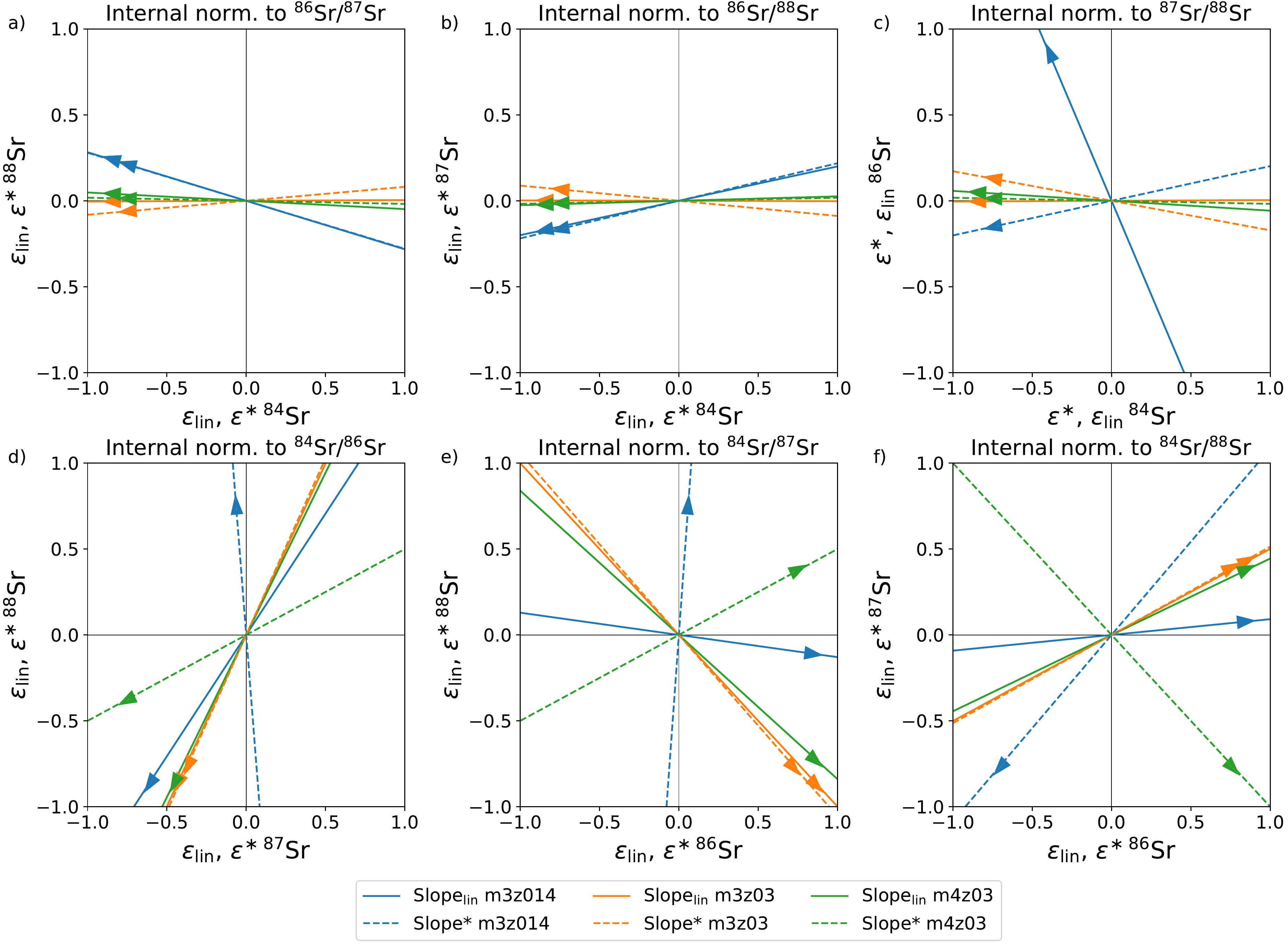}
\caption{Same as Figure~\ref{fig:Srexpl} but for the 6 possible different internal normalisation for Sr (with the corresponding internal normalisation ratio indicated at the top of each panel) applied to the three AGB models considered here (M=3 \msun with Z=0.014 and 0.03, and M=4 with Z=0.03), represented each with the different colors indicated in the legend.
}
\label{fig:Srmore}       
\end{figure}

Figure~\ref{fig:Srmore} shows all the possible normalisation for the three models considered here, where we can identify which normalisation better preserves the large difference in the relative over-production of the Sr isotopes resulting from changing the metallicity from 0.014 (blue line) to 0.03 (orange and green lines). The clearest signature of the different behaviours is shown by the \iso{87}Sr/\iso{88}Sr normalisation, where correlated variations are only appearing for the Z=0.014 model while for the other models all the lines are almost flat. Using \iso{84}Sr/\iso{87,88}Sr as normalising ratios also produce significant deviations between the different models, while with the often used normalisation ratio \iso{86}Sr/\iso{88}Sr, it may be more difficult to pick up the difference between these specific models, unless the error bars on the data were smaller than the plotted variation, e.g., less than $\sim\pm$0.10 in $\epsilon$\iso{87}Sr and for $\epsilon$\iso{84}Sr$=-1$.

\subsection{Molybdenum}
\label{sec:Mo}

Molybdenum has 7 stable isotopes, of which two, the $p$-only \iso{92}Mo and the $r$-only \iso{100}Mo are not produced in AGB stars, while all the others are produced to variable degrees (Table~\ref{tab:models}). The number of possible normalisation is 21 and for each given normalisation, 10 different plots can be produced for all the possible combinations of two isotopic ratios. All these possibilities are of interest and should be considered in a dedicated study such as that of \cite{stephan21}. Here, we will focus exclusively on 
the comparison of the model predictions with the data trends shown in Figure~\ref{fig:Mo}.
 
\begin{figure}
\centering
  \includegraphics[width=8 cm]{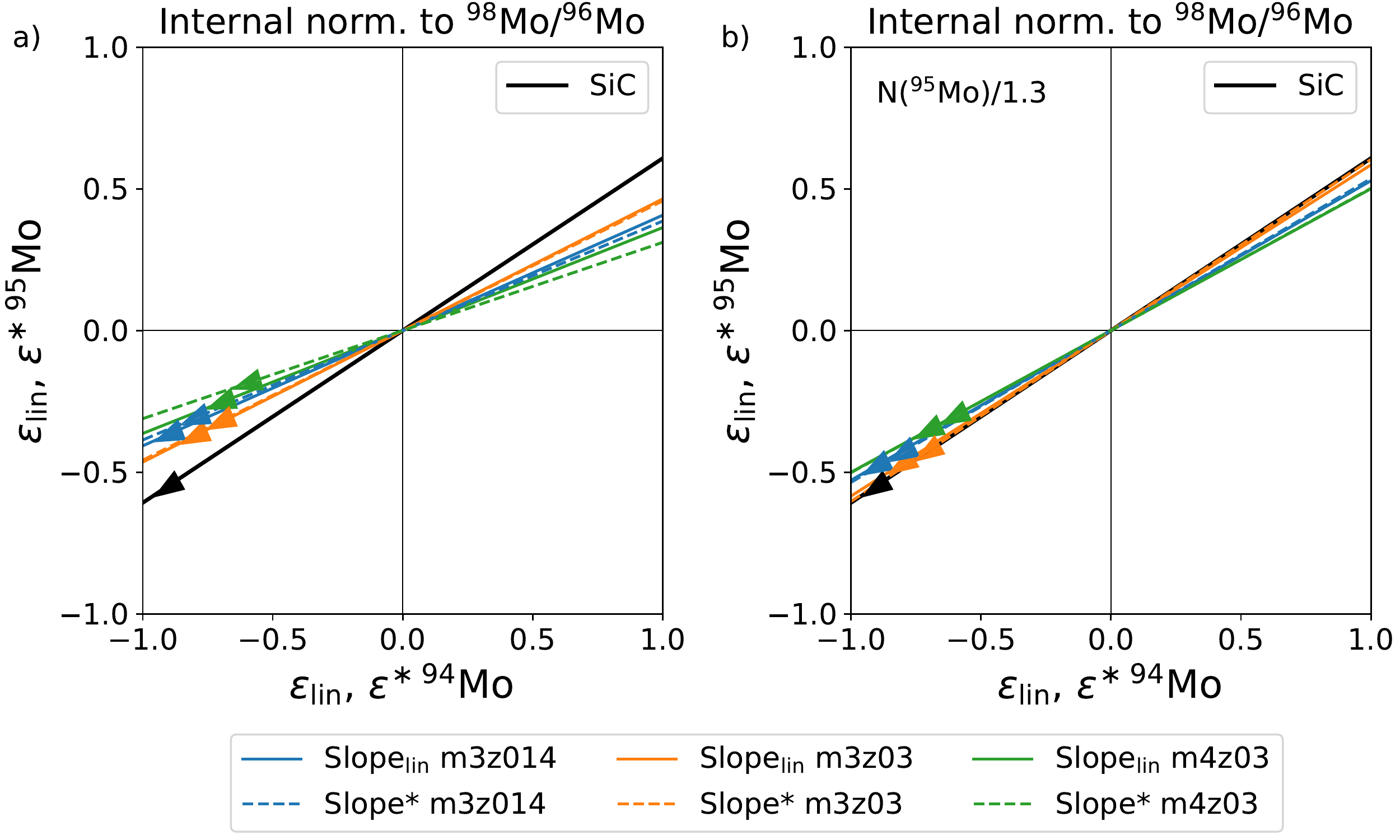}
\caption{Slopes predicted using the three different AGB models considered here (represented each with the color  indicated in the legend) for the same Mo isotopes in the same normalisation as Figure~\ref{fig:Mo}. In the right panel the abundance by number (N) of \iso{95}Mo was divided by 1.3 to mimic the effect of the correspondingly higher neutron-capture cross section reported by \cite{koehler22}. The solid black line represents the mainstream SiC trend line from Figure~\ref{fig:Mo}
.
}
\label{fig:Mopred}       
\end{figure} 
 
Differently to Sr, the models do not show large variations among each other in the relative production factors of Mo (see numbers in brackets in Table~\ref{tab:models}). They show variations of the order of 10\% only, and all in the same direction relative to the $s$-only \iso{96}Mo. This is because the Mo ratios that involve the isotopes with mass 95, 96, 97, and 98 mostly depend on the ratio of their neutron-capture reaction rates (although in some conditions the branching point at \iso{95}Zr can be marginally activated, by-passing \iso{95}Mo and \iso{96}Mo). These rates have typically a mild temperature dependence because they decrease by 10\%, 14\%, and 23\% between 5 and 30 KeV for \iso{95}Mo, \iso{97}Mo, and \iso{98}Mo, respectively. The rate of \iso{96}Mo instead decreases more, by 50\% (see kadonis.org database \cite{dillmann06}). Some minor production of the classical $p$-only isotope \iso{94}Mo may also occur during the $s$ process, due to neutron captures on the \iso{92}Mo and \iso{93}Nb initially present in the star. 

Figure~\ref{fig:Mopred} shows the model predictions for Mo in the representation as used in Figure~\ref{fig:Mo}. The slopes do not vary significantly either when the ratios are internally normalised, or when different models are considered. All the models reported here predict slopes between 0.35 and 0.45. Some models of lower mass (down to 2 \msun, not discussed here in detail) can reach a slope of 0.5. The direction of the $s$-process excess in the \iso{98}Mo/\iso{96}Mo normalisation is always towards negative values, as \iso{96}Mo is an $s$-only isotope and the most over-abundant in AGB models (Table~\ref{tab:models}). 

The slope of the MS SiC grains trend, almost parallel to the bulk meteoritic slopes, is equal to 0.608 $\pm$ 0.007 (1$\sigma$; Figure~\ref{fig:Mo}). Therefore, AGB models show a discrepancy relative to this value. However, the neutron-capture rates currently used in the models are based on experiments carried out in 1987 \cite{winters87}. A recent analysis for \iso{95}Mo \cite{koehler22} resulted in a rate with values highly discrepant from the old values, with the new rate up to 20-30\% higher than the older rate. To mimic the effect of these new rates, we divided the \iso{95}Mo abundances in the models considered here by 1.3 (as a higher rate results in a lower abundance). In this case, and specifically for the 3 \msun, Z=0.03 model, the slope become the same as that of the stardust and bulk meteoritic data. This modification of the \iso{95}Mo neutron-capture cross section was already predicted by Lugaro et al. \cite{lugaro03grains} on the basis of the SiC data and anticipated by Koehler et al.~\cite{koehler08}. This example clearly demonstrates that improvements in the estimates of neutron-capture cross sections are an essential ingredient to interpret bulk meteoritic data. 

\section{Conclusions}
\label{sec:conclusions}

We present the methodology required to transform stellar model predictions into the representation required to compare them to bulk meteorite data, which removes the need to accurately know the effect of the ubiquitous process of mass-dependent fractionation (MDF) in the data. We motivated the necessity of using such representation as the way by which stellar nucleosynthesis models can be exploited together with meteorite isotope data to shed light on the formation of the Solar System and the environment where this occurred. We have used some examples from the $s$ process as case studies to show in detail the effect of such normalisation (using Sr as the example) and to stress the fact that it is not possible to predict a slope, not even qualitatively, without accurate calculations of the effect of internal normalisation. The calculations per se are simple enough as various approximations, such as Eq.~\ref{eq:method5}, are valid and can be used to derive the mixing line with an end-member of stellar composition (see also Supplementary Material). The complications are due to the large number of possibilities to be explored for each element. In general, the choice of the isotopic ratios to use for the double normalisation needs to match that of the data, as determined by analytical limits and uncertainties. Still, careful consideration of the model predictions is also pertinent. Depending on the models and the element considered, some results may become more evident by using one normalising ratio rather than another, as shown here for the example of the $s$-process Sr (Figure~\ref{fig:Srmore}). It is possible to identify the effect of both the properties of the stellar models (e.g., mass and metallicity) and of the nuclear physics input via a careful analysis of each isotopic ratio. As illustrative examples, we have shown here that Sr is strongly affected by stellar metallicity, while Mo is mostly sensitive to the nuclear physics input. Overall, a huge amount of future work is required to investigate the significant potential of mapping stellar nucleosynthesis models onto the representation explained here. Each element, each stellar model, and each comparison to available data and their interpretation should be treated in detail, not only for AGB $s$-process predictions but also for core-collapse supernova models. Within the ChETEC-INFRA \url{www.chetec-infra.eu} project (to run between 2021 and 2025) we are providing the tools and the models to perform these tasks.


%
%

\section{Supplementary Material}
\label{sec:supplementary}
The spreadsheet \textit{NormalisationComparison.xlsx} contains a comparison between $\epsilon$ values calculated using the exponential (Eq.~\ref{eq:method4}; ``Exponential''), the linear (Eq.~\ref{eq:method5}; ``Linear'') and the linear equation in Dauphas et al. \cite{dauphas04} (``Dauphas'') for a selection of $s$-process models from the FRUITY database \cite{cristallo11FRUITY} for all elements listed in Table~\ref{tab:normalisations}. Mixing lines were created from these values by dividing the calculated offsets for two different isotopic ratios.
The ``Exponential'' values were calculated by creating a synthetic sample (see Eq.~\ref{eq:intro1}) where the largest $\epsilon$ offset is always equal to 1 or -1. The ``Linear'' and ``Dauphas'' values were calculated directly from the stellar abundances and then renormalised such that the largest offset was equal to 1 or -1, which enables direct comparison between the different methods. The calculations were performed using the solar values from Lodders et al. \cite{lodders09} (L09) and Anders \& Grevesse \cite{anders89} (AG89) and the same values also renormalised to the ratios listed in Table~\ref{tab:normalisations}. As can be seen in the figures included in each sheet, there is very little difference in the slopes calculated with the different methods and solar values.

A Jupyter notebook is available at \url{https://github.com/mattias-ek/internal\_normalisation}{}, which can generate all figures included in the paper as well as the data included in the spreadsheet discussed above.

\begin{acknowledgements}
This paper is based on work carried out as part of Work Package 9 of the ChETEC-INFRA project funded by the European Union’s Horizon 2020 research and innovation program under grant agreement No 101008324, the ERC-CoG-2016 RADIOSTAR project (Grant Agreement 724560), and the Hungarian NKFI(OTKA) K-138031 project. GVM was supported by the European Research Council (Advanced Grant Agreement 833275—DEEPTIME). IO was supported by the Carlsberg Foundation (CF20\_0209) and the Villum Fonden (00025333). MS and ME acknowledge funding from the Swiss Science Foundation.
We also acknowledge the support of the NuGrid collaboration, http://www.nugridstars.org, the Joint Institute for Nuclear Astrophysics - Center for the Evolution of the Elements, USA and the the US IReNA Accelnet network (Grant No. OISE-1927130). We thank the referees, Thomas Stephan and Larry Nittler, for their careful reading and for providing us with many comments, which have helped us to improve the manuscript. We thank Nicolas Dauphas and Timo Hopp for discussion. ML thanks Evelin B\'anyai for help with python and making figures. ML and MP are  deeply grateful for the opportunity to contribute to this volume in memory of Franz Kaeppeler, who has always been a friend and a mentor to them.

\end{acknowledgements}


%
%


\begin{thebibliography}{112}
\expandafter\ifx\csname natexlab\endcsname\relax\def\natexlab#1{#1}\fi

\bibitem{suess56}
{Suess} HE, {Urey} HC.
\newblock \textit{{\it Rev. Mod. Phys.}} 28(1):53--74 (1956)

\bibitem{seeger65}
{Seeger} PA, {Fowler} WA, {Clayton} DD.
\newblock \textit{\apjs} 11:121 (1965)

\bibitem{clayton61}
{Clayton} DD, {Fowler} WA, {Hull} TE, {Zimmerman} BA.
\newblock \textit{Annals of Physics} 12(3):331--408 (1961)

\bibitem{clayton74}
{Clayton} DD, {Ward} RA.
\newblock \textit{\apj} 193:397--400 (1974)

\bibitem{kaeppeler82}
{Kaeppeler} F, {Beer} H, {Wisshak} K, {Clayton} DD, {Macklin} RL, {Ward} RA.
\newblock \textit{\apj} 257:821--846 (1982)

\bibitem{kaeppeler93}
{Kaeppeler} F, {Schanz} W, {Wisshak} K, {Reffo} G.
\newblock \textit{\apj} 410:370 (1993)

\bibitem{kaeppeler89}
{Kappeler} F, {Beer} H, {Wisshak} K.
\newblock \textit{Reports on Progress in Physics} 52(8):945--1013 (1989)

\bibitem{raiteri91}
{Raiteri} CM, {Busso} M, {Gallino} R, {Picchio} G, {Pulone} L.
\newblock \textit{\apj} 367:228 (1991)

\bibitem{straniero95}
{Straniero} O, {Gallino} R, {Busso} M, {Chiefei} A, {Raiteri} CM, et~al.
\newblock \textit{\apjl} 440:L85--L87 (1995)

\bibitem{gallino98}
{Gallino} R, {Arlandini} C, {Busso} M, {Lugaro} M, {Travaglio} C, et~al.
\newblock \textit{\apj} 497:388 (1998)

\bibitem{arlandini99}
{Arlandini} C, {K{\"a}ppeler} F, {Wisshak} K, {Gallino} R, {Lugaro} M, et~al.
\newblock \textit{\apj} 525:886--900 (1999)

\bibitem{the07}
{The} LS, {El Eid} MF, {Meyer} BS.
\newblock \textit{\apj} 655:1058--1078 (2007)

\bibitem{pignatari10}
{Pignatari} M, {Gallino} R, {Heil} M, {Wiescher} M, {K{\"a}ppeler} F, et~al.
\newblock \textit{\apj} 710:1557--1577 (2010)

\bibitem{bisterzo10PaperI}
{Bisterzo} S, {Gallino} R, {Straniero} O, {Cristallo} S, {K{\"a}ppeler} F.
\newblock \textit{\mnras} 404:1529--1544 (2010)

\bibitem{travaglio04}
{Travaglio} C, {Hillebrandt} W, {Reinecke} M, {Thielemann} FK.
\newblock \textit{\aap} 425:1029--1040 (2004)

\bibitem{prantzos18}
{Prantzos} N, {Abia} C, {Limongi} M, {Chieffi} A, {Cristallo} S.
\newblock \textit{\mnras} 476(3):3432--3459 (2018)

\bibitem{kobayashi20}
{Kobayashi} C, {Karakas} AI, {Lugaro} M.
\newblock \textit{\apj} 900(2):179 (2020)

\bibitem{dauphas16}
{Dauphas} N, {Schauble} EA.
\newblock \textit{Annual Review of Earth and Planetary Sciences} 44:709--783
  (2016)

\bibitem{yokoyama16}
{Yokoyama} T, {Walker} RJ.
\newblock \textit{Reviews in Mineralogy and Geochemistry} 81(1):107--160 (2016)

\bibitem{qin16}
{Qin} L, {Carlson} RW.
\newblock \textit{Geochemical Journal} 50(1):43--65 (2016)

\bibitem{bermingham20}
{Bermingham} KR, {F{\"u}ri} E, {Lodders} K, {Marty} B.
\newblock \textit{\ssr} 216(8):133 (2020)

\bibitem{kleine20}
{Kleine} T, {Budde} G, {Burkhardt} C, {Kruijer} TS, {Worsham} EA, et~al.
\newblock \textit{\ssr} 216(4):55 (2020)

\bibitem{mezger20}
Mezger K, Sch{\"o}nb{\"a}chler M, Bouvier A.
\newblock \textit{Space science reviews} 216(2):1--24 (2020)

\bibitem{trinquier09}
{Trinquier} A, {Elliott} T, {Ulfbeck} D, {Coath} C, {Krot} AN, {Bizzarro} M.
\newblock \textit{Science} 324(5925):374 (2009)

\bibitem{steele12}
{Steele} RCJ, {Coath} CD, {Regelous} M, {Russell} S, {Elliott} T.
\newblock \textit{\apj} 758(1):59 (2012)

\bibitem{ek20}
{Ek} M, {Hunt} AC, {Lugaro} M, {Sch{\"o}nb{\"a}chler} M.
\newblock \textit{{\it Nature Astronomy}} 4:273--281 (2020)

\bibitem{akram15}
Akram W, Sch{\"o}nb{\"a}chler M, Bisterzo S, Gallino R.
\newblock \textit{Geochimica et Cosmochimica Acta} 165:484--500 (2015)

\bibitem{hutchison22}
Hutchison MA, Bod{\'e}nan JD, Mayer L, Sch{\"o}nb{\"a}chler M.
\newblock \textit{Monthly Notices of the Royal Astronomical Society}
  512(4):5874--5894 (2022)

\bibitem{nanne19}
{Nanne} JAM, {Nimmo} F, {Cuzzi} JN, {Kleine} T.
\newblock \textit{{\it Earth Planet. Sci. Lett.}} 511:44--54 (2019)

\bibitem{lichtenberg21}
Lichtenberg T, Drazkowska J, Schönbächler M, Golabek GJ, Hands TO.
\newblock \textit{Science} 371(6527):365--370 (2021)

\bibitem{haba21}
Haba MK, Lai YJ, Wotzlaw JF, Yamaguchi A, Lugaro M, Sch{\"o}nb{\"a}chler M.
\newblock \textit{Proceedings of the National Academy of Sciences}
  118(8):e2017750118 (2021)

\bibitem{zinner14}
{Zinner} E. 2014.
\newblock In \textit{Meteorites and Cosmochemical Processes. Vol. 1 Treatise on
  Geochemistry, 2nd Ed. Elsevier, Oxford}, eds. AM~{Davis}, HD~Exec.
  Eds.~{Holland}, TK~K.

\bibitem{nittler16}
{Nittler} LR, {Ciesla} F.
\newblock \textit{\araa} 54:53--93 (2016)

\bibitem{lugaro18grains}
{Lugaro} M, {Karakas} AI, {Pet{\H{o}}} M, {Plachy} E.
\newblock \textit{\gca} 221:6--20 (2018)

\bibitem{liu19Mo}
{Liu} N, {Stephan} T, {Cristallo} S, {Gallino} R, {Boehnke} P, et~al.
\newblock \textit{\apj} 881(1):28 (2019)

\bibitem{gallino97}
{Gallino} R, {Busso} M, {Lugaro} M. 1997.
\newblock In \textit{American Institute of Physics Conference Series}, ed.
  {E.~K.~Zinner \& T.~J.~Bernatowicz}, vol. 402 of \textit{American Institute
  of Physics Conference Series}

\bibitem{karakas14dawes}
{Karakas} AI, {Lattanzio} JC.
\newblock \textit{\pasa} 31:id. e030 (2014)

\bibitem{dauphas04}
{Dauphas} N, {Davis} AM, {Marty} B, {Reisberg} L.
\newblock \textit{{\it Earth Planet. Sci. Lett.}} 226(3-4):465--475 (2004)

\bibitem{burkhardt21}
{Burkhardt} C, {Spitzer} F, {Morbidelli} A, {Budde} G, {Render} JH, et~al.
\newblock \textit{Science Advances} 7(52):eabj7601 (2021)

\bibitem{desch18}
{Desch} SJ, {Kalyaan} A, {Alexander} CMO'D.
\newblock \textit{\apjs} 238(1):11 (2018)

\bibitem{trinquier07}
{Trinquier} A, {Birck} JL, {All{\`e}gre} CJ.
\newblock \textit{\apj} 655(2):1179--1185 (2007)

\bibitem{leya08}
Leya I, Sch{\"o}nb{\"a}chler M, Wiechert U, Kr{\"a}henb{\"u}hl U, Halliday AN.
\newblock \textit{Earth and Planetary Science Letters} 266(3-4):233--244 (2008)

\bibitem{burkhardt11}
{Burkhardt} C, {Kleine} T, {Oberli} F, {Pack} A, {Bourdon} B, {Wieler} R.
\newblock \textit{{\it Earth Planet. Sci. Lett.}} 312:390--400 (2011)

\bibitem{nittler:18}
{Nittler} LR, {Alexander} CMO'D, {Liu} N, {Wang} J.
\newblock \textit{\apjl} 856(2):L24 (2018)

\bibitem{jones:19a}
{Jones} S, {R{\"o}pke} FK, {Fryer} C, {Ruiter} AJ, {Seitenzahl} IR, et~al.
\newblock \textit{\aap} 622:A74 (2019)

\bibitem{denhartogh22}
{den Hartogh} J, {Pet{\"o}} MK, {Lawson} T, {Sieverding} A, {Brinkman} H,
  et~al.
\newblock \textit{\apj} 927(2):220 (2022)

\bibitem{kruijer17}
{Kruijer} TS, {Burkhardt} C, {Budde} G, {Kleine} T.
\newblock \textit{{\it Proc. Natl. Acad. Sci. USA}} 114(26):6712--6716 (2017)

\bibitem{budde19}
Budde G, Burkhardt C, Kleine T.
\newblock \textit{Nature Astronomy} 3(8):736--741 (2019)

\bibitem{brennecka20}
Brennecka GA, Burkhardt C, Budde G, Kruijer TS, Nimmo F, Kleine T.
\newblock \textit{Science} 370(6518):837--840 (2020)

\bibitem{stephan19}
{Stephan} T, {Trappitsch} R, {Hoppe} P, {Davis} AM, {Pellin} MJ, {Pardo} OS.
\newblock \textit{\apj} 877(2):101 (2019)

\bibitem{cowan21}
{Cowan} JJ, {Sneden} C, {Lawler} JE, {Aprahamian} A, {Wiescher} M, et~al.
\newblock \textit{Reviews of Modern Physics} 93(1):015002 (2021)

\bibitem{meyer00MoZr}
{Meyer} BS, {Clayton} DD, {The} LS.
\newblock \textit{\apjl} 540(1):L49--L52 (2000)

\bibitem{pignatari18}
{Pignatari} M, {Hoppe} P, {Trappitsch} R, {Fryer} C, {Timmes} FX, et~al.
\newblock \textit{\gca} 221:37--46 (2018)

\bibitem{liu18SN}
{Liu} N, {Nittler} LR, {Alexander} CMO'D, {Wang} J.
\newblock \textit{{\it Science Advances}} 4(1):eaao1054 (2018)

\bibitem{bisterzo11PaperII}
{Bisterzo} S, {Gallino} R, {Straniero} O, {Cristallo} S, {K{\"a}ppeler} F.
\newblock \textit{\mnras} 418:284--319 (2011)

\bibitem{stephan21}
{Stephan} T, {Davis} AM.
\newblock \textit{\apj} 909(1):8 (2021)

\bibitem{simon09}
{Simon} JI, {DePaolo} DJ, {Moynier} F.
\newblock \textit{\apj} 702(1):707--715 (2009)

\bibitem{savina03}
{Savina} MR, {Pellin} MJ, {Tripa} CE, {Veryovkin} IV, {Calaway} WF, {Davis} AM.
\newblock \textit{\gca} 67(17):3215--3225 (2003)

\bibitem{liu15SrBa}
{Liu} N, {Savina} MR, {Gallino} R, {Davis} AM, {Bisterzo} S, et~al.
\newblock \textit{\apj} 803:12 (2015)

\bibitem{trappitsch18Fe60}
{Trappitsch} R, {Boehnke} P, {Stephan} T, {Telus} M, {Savina} MR, et~al.
\newblock \textit{\apjl} 857(2):L15 (2018)

\bibitem{rehkamper12}
Rehk{\"a}mper M, Sch{\"o}nb{\"a}chler M, Andreasen R.
\newblock In \textit{Isotopic analysis: fundamentals and applications using
  ICP-MS}, eds. F~Vanhaecke, P~Degryse, chap.~10. Weinheim, Germany: Wiley-VCH,
   275--315 (2012)

\bibitem{schauble04}
Schauble EA.
\newblock \textit{Reviews in mineralogy and geochemistry} 55(1):65--111 (2004)

\bibitem{leya13}
Leya I, Masarik J.
\newblock \textit{Meteoritics \& planetary science} 48(4):665--685 (2013)

\bibitem{hunt17}
{Hunt} AC, {Ek} M, {Sch{\"o}nb{\"a}chler} M.
\newblock \textit{\gca} 216:82--95 (2017)

\bibitem{bigeleisen96}
Bigeleisen J.
\newblock \textit{Journal of the American Chemical Society} 118(15):3676--3680
  (1996)

\bibitem{schonbachler16a}
Sch{\"o}nb{\"a}chler M.
\newblock In \textit{Encyclopedia of Geochemistry: A Comprehensive Reference
  Source on the Chemistry of the Earth}, ed. WM~White. Cham: Springer
  International Publishing, \url{}{https://doi.org/10.1007/978-3-319-39193-9\_111-1} (2016)

\bibitem{schonbachler16b}
Sch{\"o}nb{\"a}chler M.
\newblock In \textit{Encyclopedia of Geochemistry: A Comprehensive Reference
  Source on the Chemistry of the Earth}, ed. WM~White. Cham: Springer
  International Publishing, \url{}{https://doi.org/10.1007/978-3-319-39193-9\_113-1} (2016)

\bibitem{russell78}
{Russell} WA, {Papanastassiou} DA, {Tombrello} TA.
\newblock \textit{\gca} 42(8):1075--1090 (1978)

\bibitem{lodders09}
{Lodders} K, {Palme} H, {Gail} HP.
\newblock \textit{Landolt-B{\"o}rnstein, New Series VI/4B, 34, Chapter 4.4.}
  (2009)

\bibitem{hopp22}
{Hopp} T, {Dauphas} N, {Spitzer} F, {Burkhardt} C, {Kleine} T.
\newblock \textit{Earth and Planetary Science Letters} 577:117245 (2022)

\bibitem{trappitsch18}
{Trappitsch} R, {Stephan} T, {Savina} MR, {Davis} AM, {Pellin} MJ, et~al.
\newblock \textit{\gca} 221:87--108 (2018)

\bibitem{cook21}
Cook DL, Meyer BS, Sch{\"o}nb{\"a}chler M.
\newblock \textit{The Astrophysical Journal} 917(2):59 (2021)

\bibitem{steller22}
Steller T, Burkhardt C, Yang C, Kleine T.
\newblock \textit{Icarus} 386:115171 (2022)

\bibitem{savage22}
Savage PS, Moynier F, Boyet M.
\newblock \textit{Icarus} 386:115172 (2022)

\bibitem{charlier19}
Charlier B, Tissot F, Dauphas N, Wilson C.
\newblock \textit{Geochimica et Cosmochimica Acta} 265:413--430 (2019)

\bibitem{barzyk07}
{Barzyk} JG, {Savina} MR, {Davis} AM, {Gallino} R, {Gyngard} F, et~al.
\newblock \textit{Meteoritics and Planetary Science} 42:1103--1119 (2007)

\bibitem{nicolussi97}
{Nicolussi} GK, {Davis} AM, {Pellin} MJ, {Lewis} RS, {Clayton} RN, {Amari} S.
\newblock \textit{Science} 277:1281--1283 (1997)

\bibitem{spitzer20}
Spitzer F, Burkhardt C, Budde G, Kruijer TS, Morbidelli A, Kleine T.
\newblock \textit{The Astrophysical Journal Letters} 898(1):L2 (2020)

\bibitem{nicolussi98}
{Nicolussi} GK, {Pellin} MJ, {Lewis} RS, {Davis} AM, {Clayton} RN, {Amari} S.
\newblock \textit{\apj} 504:492--+ (1998)

\bibitem{fischer17}
Fischer-G{\"o}dde M, Kleine T.
\newblock \textit{Nature} 541(7638):525--527 (2017)

\bibitem{savina04}
{Savina} MR, {Davis} AM, {Tripa} CE, {Pellin} MJ, {Gallino} R, et~al.
\newblock \textit{Science} 303:649--652 (2004)

\bibitem{andreasen07}
Andreasen R, Sharma M.
\newblock \textit{The Astrophysical Journal} 665(1):874 (2007)

\bibitem{liu14Ba}
{Liu} N, {Savina} MR, {Davis} AM, {Gallino} R, {Straniero} O, et~al.
\newblock \textit{\apj} 786:66 (2014)

\bibitem{bermingham16}
Bermingham K, Mezger K, Scherer E, Horan M, Carlson R, et~al.
\newblock \textit{Geochimica et cosmochimica acta} 175:282--298 (2016)

\bibitem{frossard21}
Frossard P, Guo Z, Spencer M, Boyet M, Bouvier A.
\newblock \textit{Earth and Planetary Science Letters} 566:116968 (2021)

\bibitem{yin06}
{Yin} QZ, {Lee} CTA, {Ott} U.
\newblock \textit{\apj} 647(1):676--684 (2006)

\bibitem{burkhardt15Nd}
Burkhardt C, Borg L, Brennecka G, Shollenberger Q, Dauphas N, Kleine T.
\newblock \textit{Nature} 537(7620):394--398 (2016)

\bibitem{shollenberger18}
Shollenberger QR, Render J, Brennecka GA.
\newblock \textit{Earth and Planetary Science Letters} 495:12--23 (2018)

\bibitem{avila12}
{{\'A}vila} JN, {Lugaro} M, {Ireland } TR, {Gyngard} F, {Zinner} E, et~al.
\newblock \textit{\apj} 744(1):49 (2012)

\bibitem{cristallo11FRUITY}
{Cristallo} S, {Piersanti} L, {Straniero} O, {Gallino} R, {Dom{\'{\i}}nguez} I,
  et~al.
\newblock \textit{\apjs} 197:17 (2011)

\bibitem{anders89}
{Anders} E, {Grevesse} N.
\newblock \textit{\gca} 53:197--214 (1989)

\bibitem{dauphas:10}
{Dauphas} N, {Remusat} L, {Chen} JH, {Roskosz} M, {Papanastassiou} DA, et~al.
\newblock \textit{\apj} 720(2):1577--1591 (2010)

\bibitem{amari95b}
{Amari} S, {Hoppe} P, {Zinner} E, {Lewis} RS.
\newblock \textit{Meteoritics} 30:679 (1995)

\bibitem{pignatari13grains}
{Pignatari} M, {Wiescher} M, {Timmes} FX, {de Boer} RJ, {Thielemann} FK, et~al.
\newblock \textit{\apj} 767:L22 (2013)

\bibitem{ott90}
{Ott} U, {Begemann} F.
\newblock \textit{\apjl} 353:L57 (1990)

\bibitem{hoppe97}
{Hoppe} P, {Ott} U. 1997.
\newblock In \textit{American Institute of Physics Conference Series}, ed.
  {T.~J.~Bernatowicz \& E.~Zinner}, vol. 402 of \textit{American Institute of
  Physics Conference Series}

\bibitem{liu18c13pocket}
{Liu} N, {Gallino} R, {Cristallo} S, {Bisterzo} S, {Davis} AM, et~al.
\newblock \textit{\apj} 865(2):112 (2018)

\bibitem{lugaro20}
{Lugaro} M, {Cseh} B, {Vil{\'a}gos} B, {Karakas} AI, {Ventura} P, et~al.
\newblock \textit{\apj} 898(2):96 (2020)

\bibitem{travaglio18}
{Travaglio} C, {Rauscher} T, {Heger} A, {Pignatari} M, {West} C.
\newblock \textit{\apj} 854:18 (2018)

\bibitem{farouqi09}
{Farouqi} K, {Kratz} KL, {Pfeiffer} B.
\newblock \textit{\pasa} 26:194--202 (2009)

\bibitem{bliss18}
{Bliss} J, {Arcones} A, {Qian} YZ.
\newblock \textit{\apj} 866(2):105 (2018)

\bibitem{kratz19}
{Kratz} KL, {Akram} W, {Farouqi} K, {Hallmann} O. 2019.
\newblock In \textit{Exotic Nuclei and Nuclear/particle AstroPhysics (VII).
  Physics with Small Accelerators}, vol. 2076 of \textit{American Institute of
  Physics Conference Series}

\bibitem{rauscher02}
{Rauscher} T, {Heger} A, {Hoffman} RD, {Woosley} SE.
\newblock \textit{\apj} 576:323--348 (2002)

\bibitem{asplund09}
{Asplund} M, {Grevesse} N, {Sauval} AJ, {Scott} P.
\newblock \textit{\araa} 47:481--522 (2009)

\bibitem{karakas16}
{Karakas} AI, {Lugaro} M.
\newblock \textit{\apj} 825:26 (2016)

\bibitem{cristallo15LEPP}
{Cristallo} S, {Abia} C, {Straniero} O, {Piersanti} L.
\newblock \textit{\apj} 801(1):53 (2015)

\bibitem{prantzos20}
{Prantzos} N, {Abia} C, {Cristallo} S, {Limongi} M, {Chieffi} A.
\newblock \textit{\mnras} 491(2):1832--1850 (2020)

\bibitem{koehler22}
{Koehler} PE.
\newblock \textit{\prc} 105(5):054306 (2022)

\bibitem{dillmann06}
{Dillmann} I, {Heil} M, {K{\"a}ppeler} F, {Plag} R, {Rauscher} T, {Thielemann}
  F. 2006.
\newblock In \textit{Capture Gamma-Ray Spectroscopy and Related Topics}, ed.
  {A.~Woehr \& A.~Aprahamian}, vol. 819 of \textit{American Institute of
  Physics Conference Series}

\bibitem{winters87}
{Winters} RR, {Macklin} RL.
\newblock \textit{\apj} 313:808 (1987)

\bibitem{lugaro03grains}
{Lugaro} M, {Davis} AM, {Gallino} R, {Pellin} MJ, {Straniero} O, {K{\"a}ppeler}
  F.
\newblock \textit{\apj} 593:486--508 (2003)

\bibitem{koehler08}
{Koehler} P, {Harvey} JA, {Guber} K, {Wiarda} DA. 2008.
\newblock In \textit{Nuclei in the Cosmos (NIC X)}

\end{thebibliography}

\end{document}